\documentclass[aps,prl,reprint,floatfix,superscriptaddress,nofootinbib]{revtex4-2}
\usepackage{amsmath,amssymb,graphicx,bm}
\usepackage[usenames,dvipsnames,svgnames,x11names,table]{xcolor}
\usepackage[pdfusetitle]{hyperref}
\usepackage{placeins}
\hypersetup{colorlinks=true,linkcolor=RoyalBlue,citecolor=RoyalBlue,urlcolor=RoyalBlue,pdfborder={0 0 0},breaklinks=true,pdflang={English}}

\newcommand{\prlsection}[1]{{\em {#1}.---~}}

\newcommand{\PRLrefsep}{%
  \par\vspace{0.75\baselineskip}%
  \noindent\makebox[\linewidth][c]{%
    \rule[0.5ex]{0.08\linewidth}{0.25pt}%
    \rule[0.5ex]{0.09\linewidth}{0.45pt}%
    \rule[0.5ex]{0.10\linewidth}{0.70pt}%
    \rule[0.5ex]{0.18\linewidth}{1.00pt}%
    \rule[0.5ex]{0.10\linewidth}{0.70pt}%
    \rule[0.5ex]{0.09\linewidth}{0.45pt}%
    \rule[0.5ex]{0.08\linewidth}{0.25pt}%
  }%
  \par\vspace{0.75\baselineskip}%
}
\usepackage{times}
\usepackage{orcidlink}

\begin{document}
\title{Universality and Dynamical Inequivalence in Isospectral Non-Hermitian Anderson Transitions}


\author{Aziz Hasan\orcidlink{0009-0007-0445-8237}} \thanks{Email to: azizhasan2037@gmail.com}
\affiliation{Department of Physical Sciences, IISER Kolkata, Mohanpur, West Bengal 741246, India}

\author{Anant Vijay Varma\orcidlink{0000-0002-7610-6317}}
\affiliation{Computational Physics Laboratory, Tampere University, P.O. Box 600, FI-33014 Tampere, Finland}

\author{Namit Anand\orcidlink{0000-0003-4116-4581}}
\affiliation{HPE Quantum, Emergent Machine Intelligence, HPE Labs, CA, USA}

\author{Sourin Das\orcidlink{0000-0002-8511-5709}} \thanks{Email to: sourin@iiserkol.ac.in}
\affiliation{Department of Physical Sciences, IISER Kolkata, Mohanpur, West Bengal 741246, India}

\date{\today}
\begin{abstract}
The Hatano Nelson paradigm establishes that extensive bulk nonreciprocity can destabilize Anderson localization via an imaginary gauge flux. Here, we demonstrate that extensive nonreciprocity is not a necessary ingredient: a single non-Hermitian boundary bond in a disordered one-dimensional ring suffices to drive the localization-delocalization transition. More generally, we construct an exactly isospectral family of non-Hermitian Hamiltonians that continuously interpolates between the uniform Hatano Nelson model and the single-bond limit. We show that the universal critical behavior encompassing spectral, eigenstate, and topological diagnostics is gauge invariant and governed solely by the total imaginary gauge flux, regardless of its spatial distribution. Remarkably, despite sharing identical spectra and critical exponents, different configurations within this isospectral family exhibit qualitatively distinct quantum dynamics, establishing a fundamental separation between static and dynamical universality in non-Hermitian systems. Specifically, the single boundary realization features rapid operator scrambling, oscillatory wavepacket acceleration, and a double re-entrant steady state entanglement transition. Finally, we propose an experimentally feasible realization based on multi-terminal topological transport, providing a realistic route toward observing boundary induced non Hermitian criticality and its unconventional dynamical signatures.
\end{abstract}

\maketitle
\prlsection{\textbf{Introduction}}
Anderson localization is one of the most profound manifestations of quantum interference in disordered systems. In one dimension (1D), arbitrarily weak disorder localizes all single-particle eigenstates, fragmenting the wave functions into exponentially localized orbitals, $|\psi(x)| \sim e^{-|x-x_0|/\xi}$, and suppressing transport in the thermodynamic limit \cite{PhysRev.109.1492}. Uniform bulk nonreciprocity, exemplified by the Hatano--Nelson (HN) model \cite{PhysRevLett.77.570,PhysRevB.56.8651,PhysRevB.58.8384}, fundamentally alters this paradigm. There, asymmetric hopping acts as an imaginary gauge field that introduces a directional spatial bias, modifying the envelope of a localized orbital to $|\psi_{L,R}(x)|\sim \exp{\left(-|x-x_0|/\xi \mp \gamma (x-x_0)\right)}$ \cite{PhysRevB.105.024303}. Delocalization occurs once the non-Hermitian (nH) amplification overcomes the Anderson decay. This understanding strongly suggests that extensive bulk nonreciprocity is an essential ingredient for nH \cite{PhysRevB.84.153101,PhysRevB.84.205128,Schomerus_2013,PhysRevLett.115.200402,San_Jose_2016,PhysRevLett.116.133903,PhysRevLett.118.040401,PhysRevLett.118.045701,PhysRevB.98.085116} delocalization, raising a fundamental question: how local a nH perturbation  can be while still destabilizing Anderson localization in one dimension?

In this Letter, we show that a single asymmetric bond connecting the ends of a disordered 1D ring is sufficient to induce the HN localization--delocalization (LDL) transition. While the resulting criticality is inherited from HN rather than fundamentally new, its realization through a strictly local nH link reveals an unexpected macroscopic sensitivity and fragility of Anderson localization of the systems. The underlying mechanism of this Single-Bond nH (SBN) model \cite{Li_2021, PhysRevB.102.075404, PhysRevB.111.064202} differs fundamentally from that of the HN model, a distinction that becomes highly apparent when observing the system's dynamics.

\begin{figure}
\centering
\includegraphics[width=1.0\linewidth]{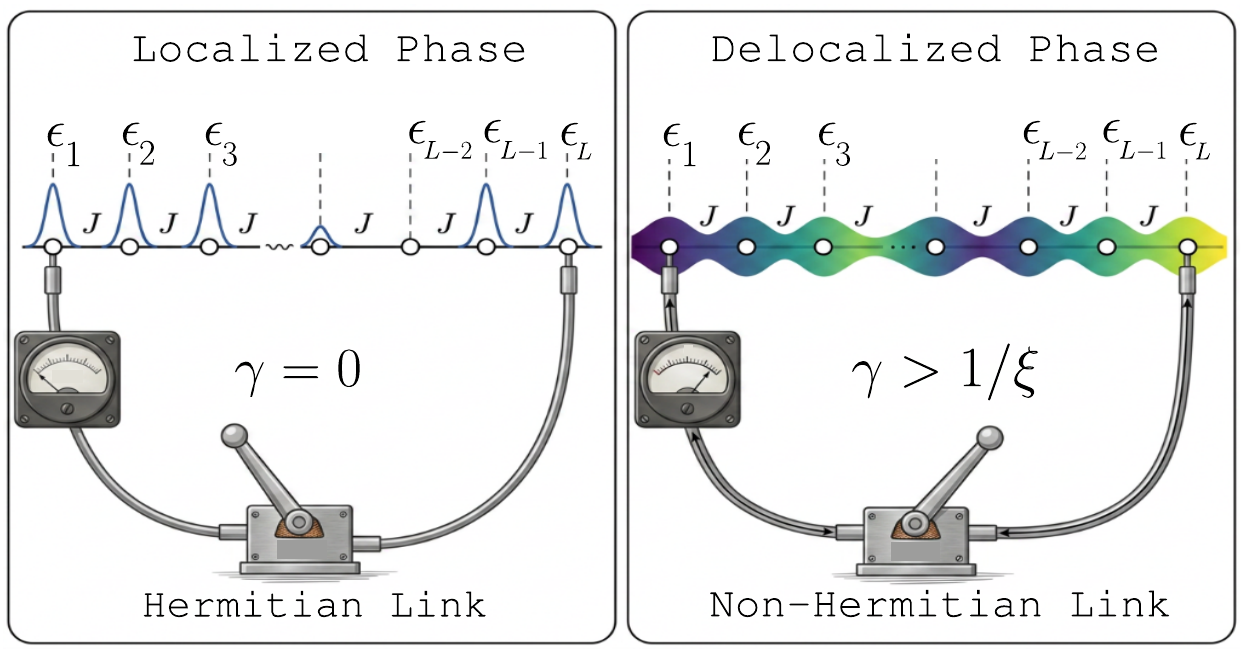}
\caption{\textit{Conceptual diagram of bulk delocalization driven by a local non-Hermitian bond.} Left: 1D lattice with random onsite potential $\epsilon_j$ and uniform coupling $J$. 
    Without non-Hermiticity, strong disorder completely localizes the wavefunctions, yielding zero current. Right: Activating a single local nH link beyond its critical threshold induces a bulk delocalization transition, resulting in a finite current. This setup effectively maps to a system pierced by an imaginary flux $\Gamma$, mimicking the delocalization mechanism of the standard HN model.}
\label{fig:figure000}
\end{figure}

To formalize these observations, we construct a continuous interpolating family of Hamiltonians that bridges the single-bond non-Hermitian (SBN) architecture and the uniform HN model under periodic boundary conditions (PBC). We demonstrate that all members of this family are mapped to one another via similarity transformations, which physically correspond to complex lattice gauge transformations. Consequently, the entire parameter space collapses into an exactly isospectral manifold uniquely characterized by a single gauge-invariant quantity: the total imaginary flux. 

It follows that the universal criticality diagnosed by standard global observables ($\mathcal{GD}$) including the fraction of complex eigenvalues ($f_c$) \cite{PhysRevLett.123.090603, PhysRevResearch.7.013098, M_k_2024, PhysRevB.56.R4333}, the inverse participation ratio ($\text{IPR}$) \cite{PhysRevLett.121.026808, PhysRevB.105.075128, PhysRevB.106.064208}, fractal dimensions ($D_n$) \cite{PhysRevB.106.094204, RevModPhys.80.1355, HENTSCHEL1983435, PhysRevLett.50.346}, state fidelity ($F$) \cite{Tu2023generalpropertiesof, PhysRevResearch.3.013015, sun2022biorthogonal, PhysRevA.98.052116}, and the topological winding number ($\mathcal{W}$) \cite{PhysRevX.8.031079, PhysRevLett.121.086803, PhysRevLett.116.133903, PhysRevX.9.041015, RevModPhys.93.015005, Ding_2022} remains strictly invariant throughout this manifold (see Supplemental Material \cite{SM} for a detailed analysis). This invariance establishes that static criticality is completely blind to the spatial distribution of nonreciprocity, setting the stage for a dramatic divergence in the system's non-equilibrium dynamics.\\

\prlsection{\textbf{Model Hamiltonian and Isospectrality}} We consider a 1D disordered ring of \(L\) sites described by
\begin{align}
\mathcal{H}(\{\alpha_j\}) = -J \sum_{j=1}^{L} \left( t_j c_j^\dagger c_{j+1} + t_j^{-1} c_{j+1}^\dagger c_j \right) + \sum_{j=1}^{L} \epsilon_j n_j,
\label{eq:hamiltonian0}
\end{align}
where \(c_j^\dagger\) (\(c_j\)) creates (annihilates) a fermion on site \(j\), \(n_j=c_j^\dagger c_j\), and PBC impose \(c_{L+1}\equiv c_1\). The asymmetric hopping amplitudes are parametrized as
\(t_j=e^{\gamma-g_j}\), with \(g_j=\alpha_{j+1}-\alpha_j\), while \(\epsilon_j\in[-w/2,w/2]\) denotes the quenched onsite disorder. The net nH gauge flux threading the ring is $\Gamma=\sum_{j=1}^{L}(\gamma-g_j)=L\gamma,$ which remains independent of the particular choice of the gauge field configuration \(\{\alpha_j\}\).\\

Remarkably, Hamiltonians with radically different microscopic distributions of non-reciprocity belong to the same isospectral manifold. This follows from the non-unitary similarity transformation $S=\exp\!\left(\sum_{j=1}^{L}\alpha_j n_j\right),$ under which the fermionic operators transform as
\begin{align}
    S\,c_j\,S^{-1}=e^{-\alpha_j}c_j, \quad S\,c_j^\dagger\,S^{-1}=e^{\alpha_j}c_j^\dagger.
\end{align}
Substituting these relations into the reference Hamiltonian, $\mathcal{H}(\{\alpha_j=0\})$ (HN Hamiltonian) generates the general model in Eq.~(\ref{eq:hamiltonian0}) while leaving the total imaginary flux \(\Gamma\) invariant. Consequently,
\begin{equation}
\mathcal{H}(\{\alpha_j\})=S\mathcal{H}(\{\alpha_j=0\})S^{-1},
\label{eq:isospectral}
\end{equation}

demonstrating that all Hamiltonians sharing the same \(\Gamma\) are related by an exact similarity transformation and therefore possess identical spectra.

Furthermore, All these models can be engineered via chiral edge modes within a multi-terminal \cite{ochkan2024non,yi2025nonhermitiandynamicsquantumanomalous} transport geometry (shown in Fig.~\ref{fig:junc_model}). Details regarding the experimental setup and the resulting conductance matrices are further elaborated in the ``Experimental Realization'' section. Alternatively these models can be engineered by post-selecting jump-free trajectories \cite{PhysRevLett.68.580,carmichael1993open,Chru_ci_ski_2022,PhysRevA.108.032214}. This is achieved via the local gain and loss jump operators define as (see SM~\cite{SM}), 
\begin{align}
    L_{j}^{l} &= \sqrt{J\sinh(\gamma-g_j)}\left(c_j - i c_{j+1}\right), \\
    L_{j}^{g} &= \sqrt{J\sinh(\gamma-g_j)}\left(c_j^{\dagger} - i c_{j+1}^{\dagger}\right),
\end{align}

\begin{figure}
\centering
\includegraphics[width=1.0\columnwidth]{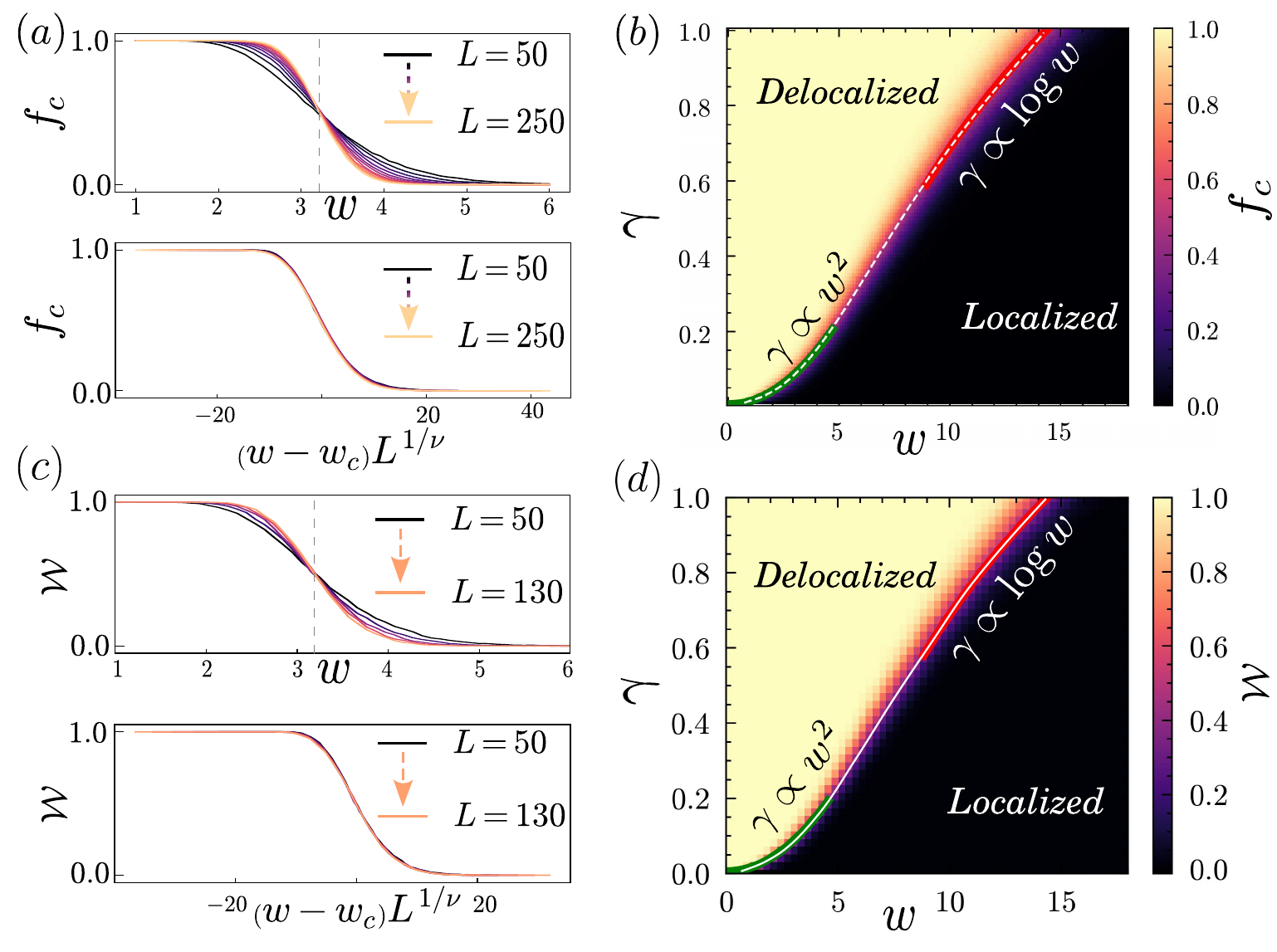}
\caption{($a$) Upper panel: $f_c$ vs.\ $w$; lower panel: $f_c$ vs.\ $(w-w_c)L^{1/\nu}$ for $\gamma=0.1$ and for $L=50,70,90\dots 250$. ($b$) $f_c$ in the $w$--$\gamma$ parameter space for $L=110$. (c) Upper panel: $\mathcal{W}$ vs.\ $w$; lower panel: $\mathcal{W}$ vs.\ $(w-w_c)L^{1/\nu}$ for $\gamma=0.1$ and $L=50,70,90,130$ ($d$) $\mathcal{W}$ in the $w$--$\gamma$ parameter space for $L=110$. All plots are made by averaging over $3000-5000$ disorder realizations.}
\label{fig:figure1}
\end{figure}

Two physically distinct limits emerge as special representatives of this manifold. Setting \(\alpha_j=0\) for all \(j\) yields the HN model, where the imaginary flux is distributed uniformly throughout the lattice, corresponding to a bond-wise flux density \(\Gamma/L=\gamma\). At the opposite extreme, choosing $\alpha_j=(j-L)\gamma$ gauges away the bulk non-reciprocity and concentrates the entire flux onto a single boundary link. This produces the SBN model, characterized by $t_{j\neq L}=1,\quad t_L=e^{L\gamma},$ thereby localizing the full imaginary flux \(\Gamma\) on the closing bond of the ring. These two models represent opposite endpoints in the space of flux distributions, yet remain exactly isospectral.

More generally, arbitrary choices of $\{\alpha_j\}$ parameterize a continuous isospectral family of Hamiltonians that smoothly interpolate between these limits while preserving the total imaginary flux. The resulting models can exhibit drastically different spatial patterns of non-reciprocity and eigenstate profiles, but their spectral properties remain identical by construction. Most importantly, the entire isospectral manifold shares an identical LDL transition. Because the diagnostic $\mathcal{GD}$ depends exclusively on the global flux $\Gamma$ rather than its microscopic spatial distribution, the critical behavior is a structural invariant of the family (see SM\cite{SM}). This establishes the total imaginary flux as the fundamental gauge-invariant quantity governing the nH delocalization transition, while the specific configuration of non-reciprocity merely selects a particular representative within the same universality class.\\

\prlsection{\textbf{Static Properties of $\bm{\mathcal{H}(\{\alpha_j\})}$ for Arbitrary $\bm{\{\alpha_j\}}$}}
\prlsection{Spectral transition} The system’s LDL phase transition is characterized by tracking $f_c$, calculated for the central $20\%$ of the spectrum, as a function of the onsite disorder strength ($w$) at fixed non-Hermiticity ($\gamma = 0.1$). The eigenenergies are defined as complex if $|Im(E)| \geq 10^{-13}$. This approach highlights the emergence of complex energies and allows precise identification of the phase boundary \cite{PhysRevLett.123.090603,PhysRevResearch.7.013098,M_k_2024,PhysRevB.56.R4333}. Varying the system size ($L$) and performing disorder averaging over $5000$ configurations reveal a critical disorder strength, $w_c \approx 3.2$ (indicated by the gray dashed line in Fig.\ref{fig:figure1}($a$)(up)), at which $f_c$ becomes independent of system size, signaling the vanishing of the finite-size scaling exponent. This behavior is further illustrated in Fig.\ref{fig:figure1}($a$)(down), which presents the scaling collapse of $f_c$ as a function of $(w-w_c)L^{1/\nu}$ with $\nu = 2.0$.\\

\prlsection{$\mathcal{W}$ Analysis}
To distinguish the topological phases and detect bulk delocalization under PBC, we monitor the global spectral flow by introducing a uniform twist phase $\Phi/L$ to each hopping link. The nH topological invariant is defined via the $\Phi$-dependent winding number $\mathcal{W}$ relative to a complex reference energy $E_B \in \mathbb{C}$:
\begin{equation}
\mathcal{W} \equiv \int_{0}^{2\pi} \frac{d\Phi}{2\pi i} \partial_{\Phi} \ln \det [\mathcal{H}(\Phi) - E_B],
\label{eq:winding_number}
\end{equation}
where the twisted Hamiltonian is given by
\begin{equation}
\mathcal{H}(\Phi) = -J \sum_{j=1}^{L} \left( t_{j,\Phi} c_j^\dagger c_{j+1} + t_{j,\Phi}^{-1} c_{j+1}^\dagger c_j \right) + \sum_{j=1}^{L} \epsilon_j n_j.
\label{eq:hamiltonian}
\end{equation}
Here, $t_{j,\Phi} = t_{j} e^{i\Phi/L}$. By plotting $\mathcal{W}$ against $w$ and its scaling collapse against $(w-w_c)L^{1/\nu}$ for $\gamma=0.1$ [upper and lower panels of Fig.~\ref{fig:figure1}($c$), respectively], we find that the transition point (and scaling exponent, $\nu$) extracted from $\mathcal{W}$ perfectly coincides with the critical value $w_c$ obtained from $f_c$.\\

In the weak-disorder (strong-disorder) regime, the phase boundary in the $w$--$\gamma$ parameter space obeys the scaling relation $\gamma \propto w_c^2$ ($\gamma \propto \log w_c$), as depicted by the green (red) lines in the $f_c$ and $\mathcal{W}$ phase diagrams [Figs.~\ref{fig:figure1}($b$) and \ref{fig:figure1}($d$), respectively]~\cite{derrida:jpa-00209867,Gliozzi_2026,PhysRev.109.1492}. The exact equations for the green and red lines, obtained via numerical fitting, are displayed in the box below. The phase boundary extracted from $f_c$ [white dashed line in Fig.~\ref{fig:figure1}($b$)] and $\mathcal{W}$ [white solid line in Fig.~\ref{fig:figure1}($d$)] is determined by tracking the $(w, \gamma)$ coordinates where these quantities exhibit scale invariance with respect to the system size $L$. A similar scaling behavior is observed for the criticality extracted from the mean fractal dimension $\langle D_2 \rangle$, albeit with modified coefficients 
(see SM~\cite{SM}).
\vspace{-0.25cm}
\begin{flushleft}
\begin{tabular}{|c|c|c|}
\hline
 & $w<4.71$ & $w>8.9$ \\
\hline
$f_c$ or $\mathcal{W}$ &
$\gamma \approx 0.005 + 0.009\,w^2$ &
$\gamma \approx -1.305 + 0.865\,\log w$ \\
\hline
$D_2$ &
$\gamma \approx 0.0126\,w^2$ &
$\gamma \approx -1.08 + 0.81\,\log w$ \\
\hline
\end{tabular}
\end{flushleft}

\vspace{0.25cm}

\prlsection{\textbf{Dynamics of SBN}}
Equivalence in $\mathcal{GD}$ does not necessarily imply microscopic 
equivalence; Hamiltonians exhibiting identical global signatures can possess 
fundamentally distinct dynamical responses. While an exhaustive analysis of the entire family is beyond the scope of this work, we focus on two representative limiting cases: the HN model, discussed in the End Matter, and the SBN model, detailed in the following section.  \\

\prlsection{Short time behavior} We prepared an initially localized wave packet and operator in the SBN system to observe their spreading in the short-time limit. Short-time limit is defined as the time required for the wave packet to complete one full revolution around the loop when $w=0$. We plot out-of-time ordered correlator (OTOC) \cite{PhysRevB.108.134305,Maldacena_2016,swingle2018unscrambling,Xu_2024,shenker2014black}, which measures the non-commutativity of the two operators evolving with time defined for nH systems as, \begin{equation*}
C_{V,W}(t)=\frac{1}{L}\sum_{j=1}^{d}\Bigg[\frac{\|VW_t|j\rangle\|^2}{\|W_t|j\rangle\|^2}-\mathrm{Re}\,\frac{\langle j|W_t^\dagger V^\dagger W_t V|j\rangle}{\|W_t|j\rangle\|\|W_tV|j\rangle\|}\Bigg],
\end{equation*} where, $W_t = U_{t}^{\dagger}W U_{t}$ denotes the time evolution generated by, $U_{t} = e^{-i H_{SBN} t}$ in the Heisenberg picture and $\left\{|j\rangle\right\}_{j=1}^{L}$ denotes  orthonormal basis of the Hilbert space with $\mathbb{I}_{H}=\sum_{j=1}^{d} |j\rangle\langle j|$. Here $V = \hat{n}_{1}$ and $W(x) = \hat{n}_{x}$ are chosen to be local number operators.\\

To obtain additional understanding of information transport, we complement OTOC with the center-of-mass (COM) trajectory. This provides a robust characterization of how the bulk of the probability density propagates through the system.
For PBC, a gauge-invariant formulation of COM at time \(t\) is defined as $
\langle x(t)\rangle
= \frac{L}{2\pi}\,
\mathrm{Im}
\!\left[
\ln\!\left(\langle \psi(t) |\, X \,| \psi(t) \rangle
\right)
\right]$, where \(X\) is the diagonal operator $X = \mathrm{diag}\!\left(1, e^{i2\pi/L}, \ldots, e^{i2\pi(L-1)/L}\right),$ which encodes the lattice positions in a manner consistent with PBC \cite{PhysRevLett.80.1800,PhysRevB.105.024303,wsmq-kmq9,manda2026crossovers,10.21468/SciPostPhys.16.5.120}.
The time-dependent state \(|\psi(t)\rangle\) is obtained as: $|\psi(t)\rangle = |\tilde\psi(t)\rangle/\|\tilde\psi(t)\|$, where $|\tilde\psi(t)\rangle \equiv e^{-i H_{SBN} t}|\psi(0)\rangle$ with \(|\psi(0)\rangle\) chosen as a wave-packet initially localized at
\textit{site-1}.
\begin{figure*}[tbh!]
    \centering
    \includegraphics[width=1.0\linewidth]{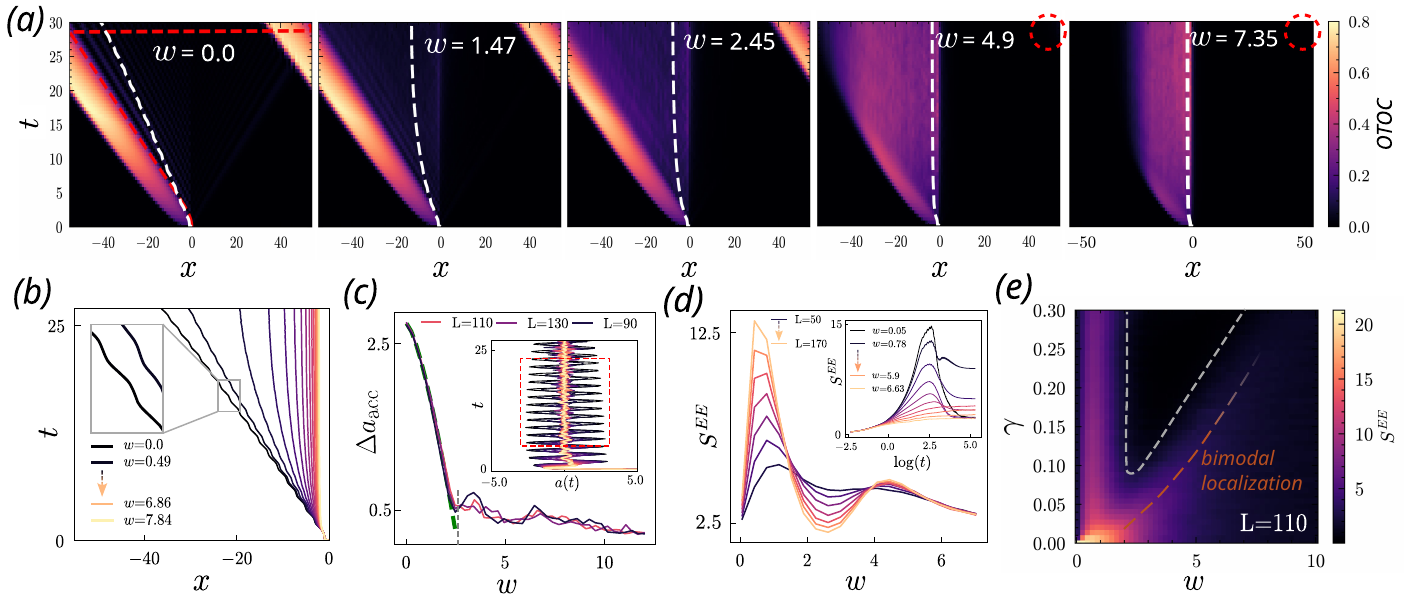}
    \caption{\textit{Dynamical and entanglement signatures of the transition for SBN}. ($a$) OTOCs for range of $w$ along with the COM trajectory (white dashed line). For $w = 0.0$, the red dashed line marks the wavefront speed.
    ($b$) COM evolution of a wave packet initially prepared at site-1 for $w\in [0.0,7.84]$, with increments of $0.49$. ($c$) Acceleration difference $\Delta a_{\mathrm{acc}}$ as a function of $w$ for system lengths $L=90, 110$, and $130$. For a given $w$, $\Delta a_{\mathrm{acc}}$ is obtained by averaging the acceleration $a(t)$ over the time interval indicated by the red dashed box in the inset, which corresponds to the period between successive encounters of the wave packet with the nonreciprocal bond and also averaging over $1000$ disorder realizations. \emph{Inset:} Time evolution of $a(t)$ from which the averages are derived. ($d$) The steady-state EE, $S^{EE}$, of the many-particle wavefunction is plotted as a function of $w$ for $L=50,70\dots 170$. The inset shows $S^{EE}$ as a function of time for $L=110$, confirming that the many particle wavefunction has reached the steady-state regime. ($e$) $S^{EE}$ in the $w$, $\gamma$ parameter space for $L=110$.}
    \label{fig:figure3}
\end{figure*}\\

Fig.~\ref{fig:figure3}($a$), $w = 0.0$ panel reveals that the operator undergoes rapid, asymmetric scrambling, effectively exploring the full Hilbert space. Consequently, this scrambling occurs at a velocity that exceeds both the wavefront propagation speed (indicated by the red dashed line) and the COM dynamics of the wavepacket (indicated by the white dashed line). At finite disorder ($w = 1.47, 2.45$) within the delocalized regime ($w < w_c$), the operator continues to ergodically sample the entire Hilbert space, even as the COM motion of the wavepacket is severely suppressed. Conversely, for strong disorder exceeding the localization threshold ($w > w_c$), this full exploration breaks down, and the operator remains confined to a restricted subspace. This bounded dynamics is visually highlighted by the red dashed circles in the 
$w = 4.9$ and $7.35$ panels of Fig.~\ref{fig:figure3}(a). Crucially, the robustness of this partial Hilbert-space exploration against strong disorder depends on the spatial proximity of the operator's support, $V$, to the nH bond. The operator spreading is constrained by a generalized Lieb-Robinson 
\cite{PhysRevLett.125.260601} speed limit, $v_{\mathrm{LR}} = 4J\cosh(L\gamma)$ (see SM \cite{SM}). 
Crucially, this bound scales with the system size $L$ as a direct consequence 
of the size-dependent nH coupling strength of SBN Hamiltonian. \\

The COM of the wavepacket exhibits an inherent oscillatory motion, as shown in Fig.~\ref{fig:figure3}($b$). This behavior can also be clearly seen in the COM acceleration, defined as $a(t)=d^2 x(t)/dt^2$, shown in the inset of Fig.~\ref{fig:figure3}($c$). The acceleration at time $t$ takes the form (see SM\cite{SM}),
\begin{align}
    a(t)\propto & \left(c_{L}^{\dagger}c_2 - c_{L-1}^{\dagger}c_{1} + h.c.\right) \nonumber\\
    &+ 2\left[1-(L-1)\cosh{L\gamma}\right]\left(n_1 - n_L\right)
\end{align}

indicating that it originates from the population imbalance between sites $1$ and $L$ and non-local boundary current term. This contribution disappears upon restoring translational invariance in the Hermitian part of the Hamiltonian (hybrid-SBN model shown in SM\cite{SM}). As $w$ increases  within the delocalized regime ($w < w_c$), 
the oscillation amplitude of the acceleration, $\Delta a_{acc} = (1/2)(a_{max} - a_{min})$ obtained via both temporal averaging (over the interval indicated by the red dashed line in the inset of 
Fig.~\ref{fig:figure3}($c$)) and ensemble averaging (over $1000$ disorder 
realizations) scales down as $2.92 - 0.7 w^{-1.4}$ 
(shown in green dashed line in Fig.~\ref{fig:figure3}(c)). Conversely, for $w > w_c$, the $\gamma$-induced acceleration is effectively suppressed by disorder. \\

\prlsection{Long time steady state behavior}
After characterizing the nH information propagation through wavepacket evolution and OTOC in the short time limit, we now turn to the steady state entanglement entropy (SSEE) of the half filled system \cite{li2023disorderinducedentanglementphasetransitions,PhysRevX.9.031009,PhysRevB.98.205136,Potter_2022,Biella_2021,PhysRevB.103.224210,Orito_2023,PhysRevLett.133.090401}. To do this, we time evolve an initially unentangled state, $|\psi(0)\rangle = \prod_{j=1}^{L/2}\hat{c}_{2j}^{\dagger}|0\rangle$ , which is a Slater determinant state. Final state after time evolution is $|\psi(t)\rangle$, having a correlation matrix $D_{i,j} = \langle\psi(t)|\hat{c}_i^{\dagger}\hat{c}_j|\psi(t) \rangle$. The Von-Neumann entanglement entropy (EE), $S_A$ between subsystem A and the rest of the system, B can be obtained by,
\begin{equation}
    S_A = - \operatorname{Tr}\left[D_A \log D_A + (1-D_A) \log (1-D_A) \right]
\end{equation}
We consider $A = \left\{L/4,L/4+1, \dots 3L/4\right\}$ and the rest is subsystem $B$. The EE can be re-written as $S_A \;=\; \sum_{\ell=1}^{|A|} h(\nu_\ell)$, where $h(\nu_\ell)= -\nu_\ell \ln \nu_\ell \;-\; (1-\nu_\ell)\ln(1-\nu_\ell)$ is a concave function of $\nu_{\ell}$ (eigenvalues of the $D_A$) with maximum at $\nu_{\ell}= 1/2$.
\begin{figure}[tbh!]
    \centering
    \includegraphics[width=1.0\linewidth]{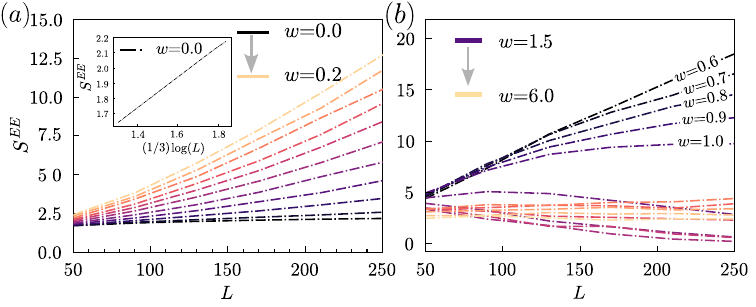}
    \caption{\label{fig:entanglement_scaling}
    \textit{$S^{\text{EE}}$ scaling for different $w$.} 
    (\textit{a}) $S^{\text{EE}}$ as a function of system size $L$ for weak disorder ($w \in [0.0, 0.2]$ in steps of $0.02$). 
    \textit{Inset:} $S^{\text{EE}}$ for the clean limit ($w=0$) compared against the conformal field theory prediction, $\frac{1}{3}\log(L)$. 
    (\textit{b}) $S^{\text{EE}}$ vs.\ $L$ for stronger disorder strengths, $w \in \{0.6, 0.7, \dots, 1.0\}$ and $\{1.5, 2.0, \dots, 6.0\}$. 
    All data are obtained by time-evolving the half-filled initial state $|\psi(0)\rangle$ up to $t=10^{6}$ (with time step $dt=0.1$), averaged over the final $10^3$ time steps and $100$ independent disorder configurations.}
    \label{fig:see_small_w}
\end{figure}
At fixed $\gamma = 0.1$, tuning the onsite disorder $w$ uncovers a re-entrant entanglement profile, where SSEE undergoes successive intervals of emergence and suppression. This intricate behavior tracks the system as it traverses four distinct phases: extended, critical, bimodally localized \cite{he2026anomalouswavepacketdynamicsonedimensional,goldberg1967computer}, and localized nH phases [Fig.~\ref{fig:figure3}($d$)]. To compute the SSEE, the system is evolved for $10^6$ time steps with a step size of $dt=0.1$ [Fig.~\ref{fig:figure3}($d$) inset], with $S^{\text{EE}}$ averaged over the final $10^3$ time steps across $1000$ disordered realizations. In the thermodynamic limit, a small onsite disorder shatters the many-particle wavefunction, preventing it from localizing near the nH bond in the SS limit. Consequently, $S^{\text{EE}}$ undergoes a crossover from logarithmic to an anomalous super-volume\cite{PRXQuantum.5.010313} law growth as shown in Fig.~\ref{fig:see_small_w}($a$). We define a super-volume law as an scaling in SSEE satisfying $S^{EE} \propto L^{\alpha}$ with $\alpha > 1$. When $w \gtrsim 0.6$ EE scales down until $w\sim 2.5$ and reemerges again [see Fig.~\ref{fig:see_small_w}($b$) and Fig.~\ref{fig:figure3}($d$)]. Notably, the second re-emergence of entanglement stems from bimodal localization and is maximally pronounced only when the nH bond is centered within the subsystem boundary.\\

\begin{figure}[tbh!]
    \centering
    \includegraphics[width=1.0\linewidth]{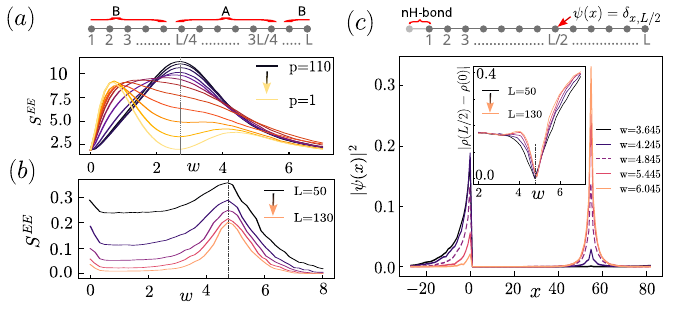}
    \caption{\textit{Entanglement and density signatures at $\gamma = 0.1$.} 
    ($a$) Many-body entanglement entropy $S_{\mathrm{EE}}$ between subsystem $A \in [L/4, 3L/4]$ and its complement $B$ for a half-filled fermionic system ($L = 110$) as a function of $w$, shown for various values of $p \in \{1, 10, 20, \dots, 110\}$. 
    ($b$) Steady-state single-particle entanglement entropy $S_{\mathrm{EE}}$ versus $w$ for the single-boundary nonreciprocal case ($p = 1$) across system sizes $L = 50, 70, \dots, 130$. 
    ($c$) Stationary spatial density profile $\left(|\psi(x)|^2 \right)$ of a single particle initially localized at site $L/2$, obtained via simultaneous disorder averaging (over $1000$ realizations) and late-time averaging, shown for selected values of $w \in [3.645, 6.045]$. 
    \emph{Inset:} The density imbalance $|\rho(L/2) - \rho(0)|$ as a function of $w$ for system lengths $L = 50, 70, \dots, 130$.}
    \label{fig:figure6a}
\end{figure}

\begin{figure}[t]
    \centering
    \includegraphics[width=1.0\linewidth]{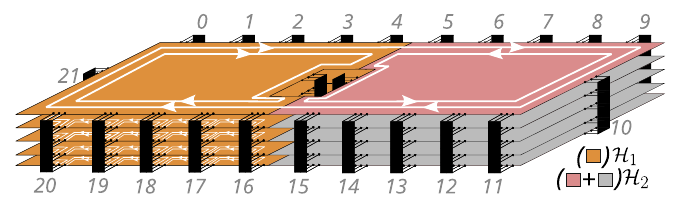}
    \caption{The Qi-Wu-Zhang model (left side of the junction, described by Hamiltonian $\mathcal{H}_1$) is coupled to the Bernevig-Hughes-Zhang model (right side of the junction, described by Hamiltonian $\mathcal{H}_2$). Black wires denote the normal leads labeled by indices $\{0, 1, \dots, 21\}$. The two intermediate leads are grounded to get only nearest-neighbor conductance in the lead basis.}
    \label{fig:junc_model}
\end{figure}

The full parameter space is mapped in Fig.~\ref{fig:figure3}($e$), revealing a pronounced valley of suppressed entanglement at large $\gamma$ (delineated by the white dashed line). Remarkably, this valley is lifted in the vicinity of the bimodal localization regime, triggering a resurgence of entanglement. The characteristic bimodal localization of a single-particle wavepacket is illustrated in Fig.~\ref{fig:figure6a}($c$). We relegate a more detailed analysis of the bimodal localization and the re-entrant emergence-suppression sequence of $S^{EE}$ to the Discussion section.\\

\vspace{0.25cm}
\prlsection{\textbf{Discussion}}
\prlsection{EE for models with $p$ bonds non Hermitian} To understand the SSEE of many particle wavefunction better we concentrate the non-Hermiticity onto $p$ consecutive bonds associated with imaginary flux $L\gamma/p$, we choose 
\begin{align}
\alpha_j=
\begin{cases}
j\left(1-\dfrac{L}{p}\right)\gamma, & 1\le j\le p,\\[4pt]
(j-L)\gamma, & p < j\le L,
\end{cases}
\end{align}
which yields an isospectral interpolation between the HN model ($p=L$) and the SBN model ($p=1$). Near the critical disorders $w\simeq w_c$, the entanglement entropy $S^{\mathrm{EE}}$ decreases smoothly as $p$ is reduced [Fig.~\ref{fig:figure6a}($a$)], demonstrating that the spatial distribution of non-Hermiticity significantly influences the entanglement structure despite leaving the spectrum unchanged.

In the strong-disorder regime ($w > w_c$), the re-entrant behavior of 
$S^{\text{EE}}$ becomes most pronounced as $p \to 1$. To elucidate the 
underlying mechanism, we track the quench dynamics of a single-particle 
wavepacket initialized at the center of the system ($x = L/2$) for SBN ($p=1$), maximizing its spatial separation from the nH bond. Upon evolving the wavefunction to its asymptotic steady-state limit, the single-particle EE develops a prominent peak at $w = 4.845$ that sharpens systematically with increasing system size $L$ [Fig.~\ref{fig:figure6a}($b$)]. To gain deeper physical insight into this anomaly, we evaluate the disorder-averaged spatial probability distribution, $\rho(x) = \langle |\langle x|\psi(t) \rangle|^2 \rangle_t$, in the steady-state limit. As shown in Fig.~\ref{fig:figure6a}($c$), a critical transition emerges near $w = 4.845$, where the wavepacket exhibits a nearly equal probability of being detected at either the nH bond (site $1$) or its initial position ($L/2$). This stark real-space crossover marks the onset of the \textit{bimodal localization} regime. For weak disorder ($w < 4.845$), the dynamics are dominated by non-reciprocal transport, preferentially localizing the wavepacket near the nH bond. Conversely, in the strong-disorder limit ($w > 4.845$), Anderson-like localization suppresses tunneling, effectively freezing the wavepacket around its initial position. This real-space crossover is quantitatively corroborated in the inset of Fig.~\ref{fig:figure6a}($c$), where the probability density contrast between sites $L/2$ and $1$ vanishes precisely at $w = 4.845$. All steady-state observables are extracted by evolving the system for $10^6$ time steps with a step size $\mathrm{d}t = 0.1$, where time averages are computed over the final $10^3$ steps and subsequently averaged over $500-1000$ independent disorder realizations. \\

Finally, it is worth noting that a growing body of work has recently begun to explore boundary perturbations in non-Hermitian systems \cite{Li_2021, PhysRevB.102.075404, PhysRevB.111.064202, PhysRevA.104.022215, PhysRevResearch.5.033058}. These previous investigations have shown that a single impurity can induce unconventional localization phenomena, serve as a sensitive probe of bulk topology, or even trigger boundary-driven many-body transitions in open quantum systems. Collectively, these works have significantly advanced our understanding of how local perturbations can reveal or modify non-Hermitian physics.\\

The central question addressed here, however, is fundamentally different. Rather than asking how a nH bond modifies the surrounding eigenstates or diagnoses an underlying bulk phase, we ask whether a strictly local nH bond can itself generate the complete macroscopic Anderson LDL transition traditionally associated with extensive bulk nonreciprocity. Remarkably, we show that the answer is affirmative.
More importantly, our analysis reveals that the transition is not controlled by the spatial extent of the non-Hermiticity but by a single gauge invariant quantity the total imaginary gauge flux. By constructing an exactly isospectral family that continuously interpolates between the HN and SBN, we demonstrate that these seemingly disparate Hamiltonians belong to the same static universality class. This establishes a unifying principle that extends well beyond the specific single-bond realization considered here.\\

Our work further uncovers a feature that has not been recognized in earlier impurity-based studies. Although all members of the isospectral family possess identical spectra, localization transitions, and topological invariants, they exhibit strikingly different quantum dynamics, including operator scrambling, wave-packet transport, and steady-state entanglement. These results demonstrate that spectral equivalence does not imply dynamical equivalence and establish a fundamental distinction between static universality, governed solely by the total imaginary gauge flux, and dynamical universality, which remains sensitive to the microscopic spatial distribution of nonreciprocity.
From this perspective, the present work is not merely another realization of the HN model with localized asymmetry, nor simply another example of impurity-induced localization. Rather, it identifies a broader organizing principle for nH disordered systems: while global critical behavior is dictated entirely by the total imaginary gauge flux, the microscopic distribution of that flux determines the non-equilibrium dynamics. This separation between universal criticality and non-universal dynamics constitutes the principal conceptual advance of our work. In conclusion, our work establishes a new organizing principle for nH disordered systems by separating universal criticality from non-universal dynamics. While previous studies focused on how local impurities modify or probe existing nH phases, we address a fundamentally different question: can a strictly local nH bond generate a macroscopic LDL transition? We show that it can.\\

\prlsection{\textbf{Experimental Realization}} Non-Hermitian physics, fundamentally governed by the interplay of gain and loss, requires precise experimental engineering to probe non-trivial topological phases. While iso-spectral Hamiltonians can be synthesized via jump-free quantum trajectories, we propose an alternative strategy that harnesses the intrinsic non-reciprocity of topological edge transport. Specifically, we demonstrate that the chiral edge modes of a Chern insulator offer a natural platform to emulate non-Hermitian Hamiltonians within a multi-terminal transport geometry, enabling the direct experimental realization and characterization of non-Hermitian topological phenomena.

We consider spinful fermions on a square lattice described by the four-component spinor $\Psi_i = (c_{iA\uparrow}, c_{iB\uparrow}, c_{iA\downarrow}, c_{iB\downarrow})^T$ at site $i$, where $A, B$ and $\uparrow, \downarrow$ denote the orbital and spin degrees of freedom, respectively. The Pauli matrices acting on the spin and orbital spaces are designated by $s_{\mu}$ and $\sigma_{\mu}$ ($\mu \in \{x,y,z\}$), with $I_s$ representing the identity matrix in spin space. The tight-binding Hamiltonian encompassing both the Qi-Wu-Zhang (QWZ) \cite{PhysRevB.74.085308} and Bernevig-Hughes-Zhang (BHZ) \cite{Bernevig_2006} models takes the general form
\begin{equation}
    H(M) = \sum_{i,\bm{\eta}} \left( \Psi_i^\dagger T_{\bm{\eta}}\Psi_{i+\bm{\eta}} + \text{H.c.} \right) + \sum_i \Psi_i^\dagger \Gamma \Psi_i,
\end{equation}
where $\bm{\eta} \in \{\hat{x}, \hat{y}\}$ denotes the nearest-neighbor vectors.\\

\textit{1. QWZ Subsystem.}---Under broken time-reversal symmetry (TRS), the system supports a higher topological phase with a total Chern number $C = 2$. For the corresponding QWZ Hamiltonian $H_{\text{QWZ}}$, the local potential $\Gamma$ and hopping matrices $T_{\bm{\eta}}$ are defined as
 \begin{align}
    \Gamma &= (2-M)(I_s \otimes \sigma_z),\quad T_x = -\frac{1}{2}(I_s \otimes \sigma_z) - \frac{i}{2}(I_s \otimes \sigma_x), \nonumber \\
    T_y    &= -\frac{1}{2}(I_s \otimes \sigma_z) - \frac{i}{2}(I_s \otimes \sigma_y).
\end{align} \\

\textit{2. BHZ Subsystem.}---When TRS is preserved, the cancellation of opposing Chern numbers in the spin-up and spin-down channels yields a net $C = 0$, driving the system into a quantum spin Hall phase hosted by helical edge states. For $H_{\text{BHZ}}$, spin-orbit coupling is introduced via $s_z$ along the $\hat{x}$-direction:
 \begin{align}
    \Gamma &= (2-M) (I_s \otimes \sigma_z),\quad T_x    = -\frac{1}{2}(I_s \otimes \sigma_z) - \frac{i}{2}(s_z \otimes \sigma_x), \nonumber \\
    T_y    &= -\frac{1}{2}(I_s \otimes \sigma_z) - \frac{i}{2}(I_s \otimes \sigma_y).
\end{align}

\begin{figure}[t]
    \centering
    \includegraphics[width=1.0\linewidth]{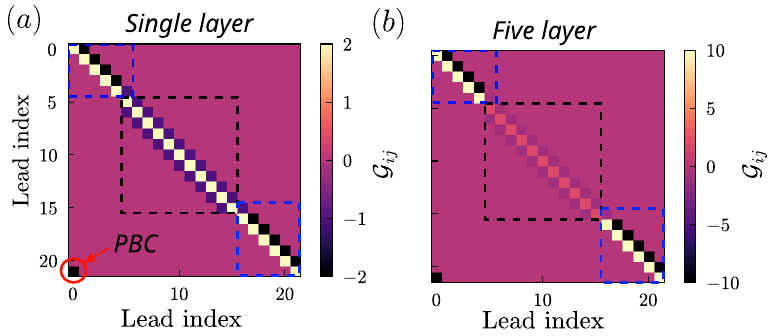}
    \caption{Conductance matrix of the multi-terminal junction for ($a$) a single layer and ($b$) $l_a = 5$ layers with vanishing inter-layer tunneling. The black and blue dashed boxes highlight the blocks corresponding to the Hermitian and non-Hermitian components of the transport Hamiltonian, respectively.}
    \label{fig:conductance_matrix}
\end{figure}

We consider a weakly coupled multi-layer heterostructure formed by joining 2D QWZ layers with 2D BHZ layers along the $x$-axis via the corresponding $T_x$ (of QWZ) hopping matrices (see Fig.~\ref{fig:junc_model}). Assuming negligible inter-layer tunneling, the total Hamiltonian of the stack is constructed as the direct sum of $l_a$ independent 2D layers:
\begin{equation}
\mathcal{H}_1 = \bigoplus_{n=1}^{l_a} H_{\text{QWZ}}(M_n), \quad \mathcal{H}_2 = \bigoplus_{n=1}^{l_a} H_{\text{BHZ}}(M_n),
\end{equation}
where $n \in \{1, \dots, l_a\}$ is the layer index. To selectively isolate and probe the topological edge states, we spatially modulate the mass parameters across the stack. In the QWZ ($\mathcal{H}_1$) subsystem, we impose a uniform topological profile $M_n = 1$. Conversely, in the BHZ ($\mathcal{H}_2$) subsystem, we break layer-translation symmetry by setting $M_1 = 1$ for the target boundary layer, while driving the remaining layers into the trivial regime via $M_{n \ge 2} \gg 4$ or $M_{n \ge 2} \ll 0$.\\

In the zero-temperature, zero-bias limit, multi-terminal quantum transport is governed by the Landauer-Büttiker \cite{PhysRevLett.57.1761} formula, where the conductance matrix elements are given by
\begin{equation}
\mathcal{G}_{ij} = \delta_{ij} N_j - \text{tr}\left( S_{ij} S_{ij}^\dagger \right).
\end{equation}
Here, $N_j$ is the number of propagating modes in lead $j$, and $S_{ij}$ is the block of the scattering matrix containing transmission amplitudes from lead $j$ to lead $i$.\\

We numerically evaluate the transport properties using the \texttt{KWANT} software package~\cite{Groth_2014} in natural units ($e = h = 1$). As illustrated in Fig.\ref{fig:junc_model}, the leads connected along the longitudinal and transverse directions of the junction are modeled as semi-infinite 1D metallic chains described by the tight-binding Hamiltonian $H_{\text{lead}} = - \sum_{i} \Psi_{i}^{\dagger} \Psi_{i+\bm{\eta}}$, where $\bm{\eta} = \hat{x}$ ($\hat{y}$) denotes the unit vector for leads connected along the transverse (longitudinal) direction. For an $l_a$-layer system, the total lead Hamiltonian is constructed via the direct sum $\mathcal{H}_{\text{lead}} = \bigoplus_{n=1}^{l_a} H_{\text{lead}}$.\\

Crucially, the effective non-Hermiticity scales monotonically as $2l_a$ with the layer index $l_a$, embedding a robust and highly controllable non-Hermitian topology within the multi-terminal conductance matrix [see Figs.~\ref{fig:conductance_matrix}($a$) and ($b$) for single-layer and 5-layer coupled QWZ-BHZ junctions, respectively]. By shifting the junction position and tuning the layer number under negligible inter-layer coupling, one can systematically realize the desired target model. Furthermore, the effective system size in the lead basis can be naturally scaled up by increasing the number of attached leads.\\

\prlsection{\textbf{Summary}} We demonstrate that concentrating nonreciprocity onto a single bond in a disordered ring can drive a sharp disorder tuned spectral and eigenstate transition under PBC. 
Static diagnostics confirm the transition and, by an explicit similarity transformation, they coincide with those of the periodic HN model. We then show that isospectrality does not imply equivalent dynamics: the single bond representative exhibits faster information spreading, quantified by a normalized non-unitary OTOC, yet slower transport of probability, quantified by the center of mass motion of a wavepacket. For free fermions at half filling, the steady state bipartite entanglement displays a pronounced suppression at intermediate disorder and revives at stronger disorder, forming a clear valley in EE in the parameter space. These results establish that the spatial placement of nonreciprocity, mediated by wavefunction structure near the special bond, can control observable non-equilibrium behavior even when static signatures are identical to models of the isospectral class.\\

\prlsection{\textbf{Acknowledgment}} A.H. wants to thank Moshe Goldstein for insightful discussions during the 'Quantum Matter in Low Dimensions' conference at IIT Gandhinagar. A.H. acknowledges University Grants Commission, India, for support in the form of a fellowship. S.D. would like to acknowledge the financial support from Anusandhan National Research Foundation (ANRF) under the MATRICS scheme (Grant No. [ANRF/ARGM/2025/002511/TS)]);  Ministry of Education, Government of India under the SPARC program (Project Code: [SPARC/2025-2026/P4086]) and National Quantum Mission under Quantum Algorithms Technical Group (TPN No.: 136428). The research of S.D. is supported partly by the International Center for Theoretical Sciences (ICTS), Bengaluru  through their Associateship Programmes. Numerical calculations were performed using the Kepler workstations at IISER Kolkata.

\bibliography{bibliography}

\section*{End Matter}
As previously established, the $\mathcal{GD}$ remain identical across the entire isospectral family. To complement this finding, we include a comparative plot illustrating the dynamics of the HN model alongside those of the SBN model, explicitly highlighting the key dynamical differences between the two.\\

\begin{figure}[tbh!]
    \includegraphics[width=1.0\linewidth]{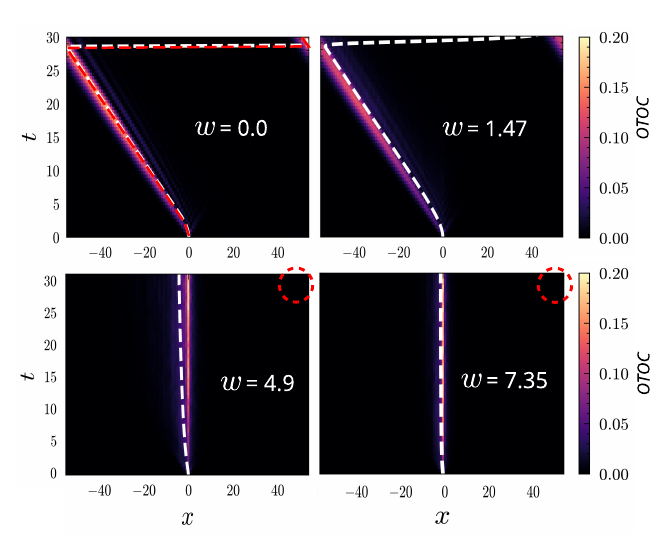}
    \caption{OTOCs for different values of $w$, with the white dashed line indicating the COM trajectory. For $w = 0.0$, the red dashed line marks the wavefront velocity.}
    \label{fig:figure55_half}
\end{figure}
Starting with OTOC profile in $w=0$ shown in Fig.\ref{fig:figure55_half} shows that, unlike the SBN model, both the COM and the wavefront in the HN model propagate with nearly the same velocity as the operator front itself. As $w$ increases, the operator continues to explore the full Hilbert space for $w < w_c$. In contrast, for $w > w_c$, the operator spreading becomes significantly suppressed, as highlighted by the red dashed circles in the $w=4.9$ and $w=7.35$ panels. \\

\begin{figure}[tbh!]
    \includegraphics[width=1.0\linewidth]{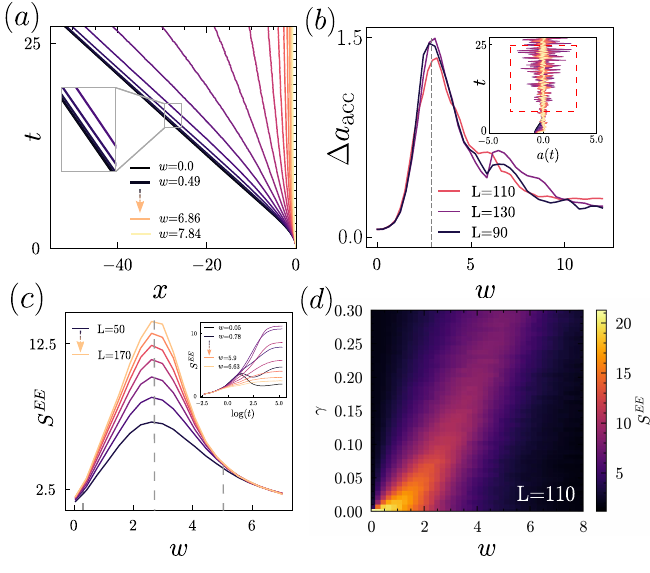}
    \caption{($a$) Time evolution of the COM for a wave packet initially localized at site $1$, shown for $w \in [0.0, 7.84]$ with increments of $0.49$. ($b$) $\Delta a_{\mathrm{acc}}$ as a function of $w$ for system sizes $L = 90$, $110$, and $130$. For each value of $w$, $\Delta a_{\mathrm{acc}}$ is obtained by first averaging the instantaneous acceleration $a(t)$ over the time window highlighted by the red dashed box in the inset and then averaging over $1000$ disorder realizations. \emph{Inset:} Time evolution of $a(t)$, from which the time-averaged values are extracted. ($c$) Steady-state entanglement entropy $S^{\mathrm{EE}}$ of the many-particle wavefunction as a function of $w$ for system sizes $L = 50, 70, \ldots, 170$. \emph{Inset:} Time evolution of $S^{\mathrm{EE}}$, demonstrating convergence to the steady-state regime. ($d$) Steady-state EE $S^{\mathrm{EE}}$ in the $(w,\gamma)$ parameter space for $L = 110$.}
    \label{fig:figure55}
\end{figure}

At $w \to 0$, the acceleration of the wavepacket COM exhibits no 
intrinsic oscillatory behavior [Fig.\ref{fig:figure55}($a,b$)], standing in 
stark contrast to the SBN model. Furthermore, the wavepacket reaches its maximum 
oscillation in acceleration near the critical regime ($w \approx w_c$). This peak occurs 
because the disorder-induced localization governed by $w$ becomes comparable to 
the chiral drift generated by $\gamma$ [Fig.\ref{fig:figure55}($b$)] at this criticality. Wavepacket traversing the disordered landscape experience a resonant maximum in back-and-forth scattering near criticality.\\

In the absence of disorder, we show that the long-time EE of the half-filled fermionic wavefunction, $|\psi(t)\rangle$, scales logarithmically with the system size as $S_{L/2} = \frac{1}{3}\log L$, mirroring the behavior of the SBN model. Within the weak disorder regime ($w \in [0.0, 1.0]$ with steps of $0.2$), unlike SBN the EE of HN consistently maintains this logarithmic scaling (see SM\cite{SM}). However, at larger disorder strengths $w$, this sub-extensive entanglement growth is disrupted by Anderson localization [Fig.~\ref{fig:figure55}($c$,$d$)]. This induces a transition from logarithmic scaling to area-law entanglement behavior, in agreement with the results reported by Li \textit{et al.}~\cite{li2023disorderinducedentanglementphasetransitions}.

\vfill\eject

\PRLrefsep

\onecolumngrid
\begin{center}
    \textbf{\large{Supplementary Material For \,\, ``Universality and Dynamical Inequivalence in Isospectral Non-Hermitian Anderson Transitions''}}
\end{center}
\vspace{0.75cm}

\twocolumngrid
\subsection{Isospectral Hamiltonians}
One can define a family of hamiltonians iso-spectral to $H_{HN}$,
\begin{align}
    \mathcal{H}(H_{HN}) &=\left\{S H_{HN} S^{-1}\;\middle|\;S \in GL(L,\mathbb{C})\right\}.\\
    \text{where,} &\quad GL(L,\mathbb{C}) = \{\, S \in \mathbb{C}^{L\times L} \mid \det S \neq 0 \,\} \nonumber
\end{align}
denotes the general linear group of degree $L$. \\

Similarity transformation can redistribute non-hermiticity  keeping the total imaginary flux same. With, $S = diag\left(e^{\alpha_1}, e^{\alpha_2}, e^{\alpha_3}, \dots e^{\alpha_L} \right) = \exp\left({\sum_{j} \alpha_{j}n_j}\right)$ total iso-spectral class can be written as
\begin{align}
    H_{HN} &= -J \sum_{j=1}^{L}\left(e^{\gamma} c_{j}^{\dagger}c_{j+1} + e^{-\gamma} c_{j+1}^{\dagger}c_{j} \right) + \sum_{j=1}^{L} \epsilon_{j} n_j
\end{align}
We now note the commutation relations $[n_{i},c^{\dagger}_{j}] = \delta_{ij}  c^{\dagger}_{i}, \,\, [n_{i},c_{j}]=-\delta_{ij}  c_{i}$ and using Baker-Campbell-Hausdorff (BCH) formalism, one can get
\begin{align}
    S c_j S^{-1} = e^{-\alpha_j} c_j, \quad S c_j^{\dagger} S^{-1} = e^{\alpha_j} c_j^{\dagger}
\end{align}
So the Hamiltonian iso-spectral to HN is,
\begin{align}
    \mathcal{H} &= S H_{HN} S^{-1}\nonumber\\
    & = -J \sum_{j=1}^{L}\left(e^{\gamma - g_{j}} c_{j}^{\dagger}c_{j+1} + e^{-\gamma + g_{j}} c_{j+1}^{\dagger}c_{j} \right) + \sum_{j=1}^{L} \epsilon_{j} n_j
    \label{eq:sm_eq1}
\end{align}

where, $g_j = \alpha_{j+1}-\alpha_{j}$ and PBC ensures $\alpha_{L+1}=\alpha_{1}$, $c_{L+1} = c_{1}$, $c_{L+1}^{\dagger} = c_{1}^{\dagger}$ and $\sum_{j=1}^{L}g_j = 0$.\\

With, $\alpha_j = (j-L)\gamma$ we get a Hamiltonian,
\begin{align}
    H_{SBN} &= -J \sum_{j=1}^{L-1}\left( c_{j}^{\dagger}c_{j+1} +h.c. \right)\nonumber\\
    & + \left(e^{L\gamma} c_L^{\dagger}c_1 + e^{-L\gamma} c_1^{\dagger}c_L \right) + \sum_{j=1}^{L} \epsilon_{j} n_j 
\end{align} also referred as SBN model in the main text. And also $H_{HN}$ referred as $\mathcal{H}\left({\alpha_j = 0}\right)$. All the Hamiltonians in the iso-spectral class have energy $\mathcal{E}=-2J\cos{(k-i \gamma)}$ for $w=0$.\\

\subsection{Identical winding number $\mathcal{W}$}
To demonstrate the invariance of the winding number $W$ within the isospectral family $\mathcal{H}$, we examine the transformation of the flux-dependent Hamiltonian under $S \in GL(L,\mathbb{C})$.\\ 

In the periodic Hatano-Nelson model, the parameter $\Phi$ introduces a magnetic flux across the boundary. The diagonal similarity transformation $S = \exp\left({\sum_{j} \alpha_{j}n_j}\right)$ scales individual hopping amplitudes and redistributes the non-Hermiticity locally. However, it preserves the global topology of the periodic system. Consequently, any member of the similarity class satisfies:
\begin{equation}
\mathcal{H}(\Phi) = S H_{\mathrm{HN}}(\Phi) S^{-1},
\end{equation}
where, 
\begin{align}
\quad H_{HN}(\Phi) = &-J \sum_{j=1}^{L}\left(e^{\gamma + i \Phi/L} c_{j}^{\dagger}c_{j+1}  + e^{-\gamma - i \Phi/L} c_{j+1}^{\dagger}c_{j} \right)\nonumber \\
 &+ \sum_{j=1}^{L} h_{j} c_j^{\dagger} c_j
\end{align}

To compute $\mathcal{W}$ for the transformed system, we evaluate the determinant of $\mathcal{H}(\Phi)$ using the multiplicative property of determinants:
\begin{align}
    \det \mathcal{H}(\Phi) &= \det \left( S H_{\mathrm{HN}}(\Phi) S^{-1} \right) \nonumber \\
    &= \det(S) \cdot \det H_{\mathrm{HN}}(\Phi) \cdot \det(S^{-1}).
\end{align}
Since $\det(S^{-1}) = \frac{1}{\det(S)}$ and $\det(S) \neq 0$, the similarity matrix contributions strictly cancel out, leaving the determinant invariant:
\begin{equation}
    \det \mathcal{H}(\Phi) = \det H_{\mathrm{HN}}(\Phi).
\end{equation}

Taking the natural logarithm and the partial derivative with respect to the flux $\Phi$ on both sides yields:
\begin{equation}
    \partial_{\Phi} \ln \det \mathcal{H} (\Phi) = \partial_{\Phi} \ln \det H_{\mathrm{HN}}(\Phi).
\end{equation}

Finally, substituting this result into the definition of the winding number gives:
\begin{align}
    \mathcal{W} &= \int_{0}^{2\pi} \frac{d\Phi}{2\pi i} \partial_{\Phi} \ln \det \mathcal{H}(\Phi)\nonumber\\
    &= \int_{0}^{2\pi} \frac{d\Phi}{2\pi i} \partial_{\Phi} \ln \det H_{\mathrm{HN}}(\Phi) = \mathcal{W}_{\mathrm{HN}}.
\end{align}

Thus, we conclude that while the similarity transformation redistributes non-Hermiticity, it leaves the global topological invariant completely unaltered.

\subsection{Identical IPR and $D_2$}
$\mathcal{H}$ is the similarity transformed Hamiltonian of $H_{HN}$ connected by $S$ matrix. Let $\bigl\{ |\psi_{L}^{n}\rangle, |\phi_{L}^{n}\rangle \bigr\}$ and $\bigl\{ |\psi_{R}^{n}\rangle, |\phi_{R}^{n}\rangle \bigr\}$ denote the $n$-th left and right eigenstates respectively of the Hamiltonians $\left\{\mathcal{H}, H\right\}$, corresponding to the eigenvalue $E_n$. Taking $\mathcal{H}$ as an example, these states satisfy the eigenvalue equations
\begin{equation}
    \mathcal{H} |\psi_{R}^{n}\rangle = E_n |\psi_{R}^{n}\rangle, \quad \langle\psi_{L}^{n}| \mathcal{H} = E_n \langle\psi_{L}^{n}|,
\end{equation}
along with the bi-orthonormality condition $\langle\psi_{L}^{n}|\psi_{R}^{m}\rangle = \delta_{n,m}$.\\

The states $|\phi_{L/R}^{n}\rangle$ satisfy identical relations with respect to $H$.
\begin{align}
&H_{HN} \lvert \phi_{R}^{n} \rangle = E_n \lvert \phi_{R}^{n}\nonumber \rangle\\
&S H_{HN} \lvert \phi_{R}^{n} \rangle = E_n S\lvert\phi_{R}^{n}\nonumber \rangle\\
&\mathcal{H} S\lvert \phi_{R}^{n} \rangle = E_n S \lvert \phi_{R}^{n}\nonumber \rangle 
\,\, \left[\because \mathcal{H} = S H_{HN} S^{-1}\nonumber\right]\\
&\mathcal{H}\lvert \psi_{R}^{n} \rangle = E_n \lvert \psi_{R}^{n} \rangle\quad \text{where,}\,\, \lvert \psi_{R}^{n}\rangle = S \lvert \phi_{R}^{n}\rangle
\end{align}
and similarly, $\langle\psi_{L}^{n}\rvert = \langle\phi_{L}^{n}\rvert S^{-1} $\\

\begin{figure}
    \centering
    \includegraphics[width=1.0\linewidth]{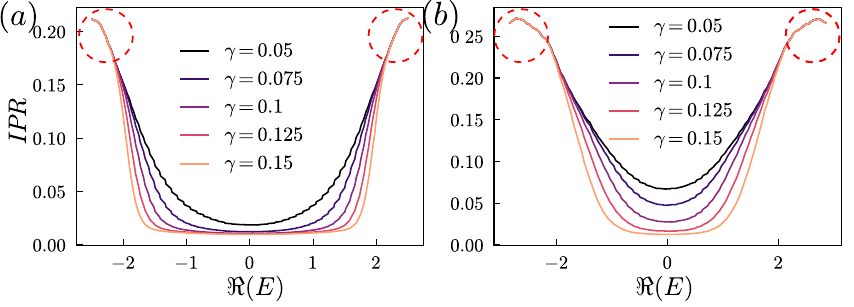}
    \caption{ $IPR$ vs $\Re(E)$ plot for different non-reciprocity parameter ($\gamma=0.1$) for ($a$) $w=2.0$ and ($b$) $w=3.0$}
    \label{fig:ipr_fidel}
\end{figure}

\begin{figure}[tbh!]
    \centering
    \includegraphics[width=1.0\linewidth]{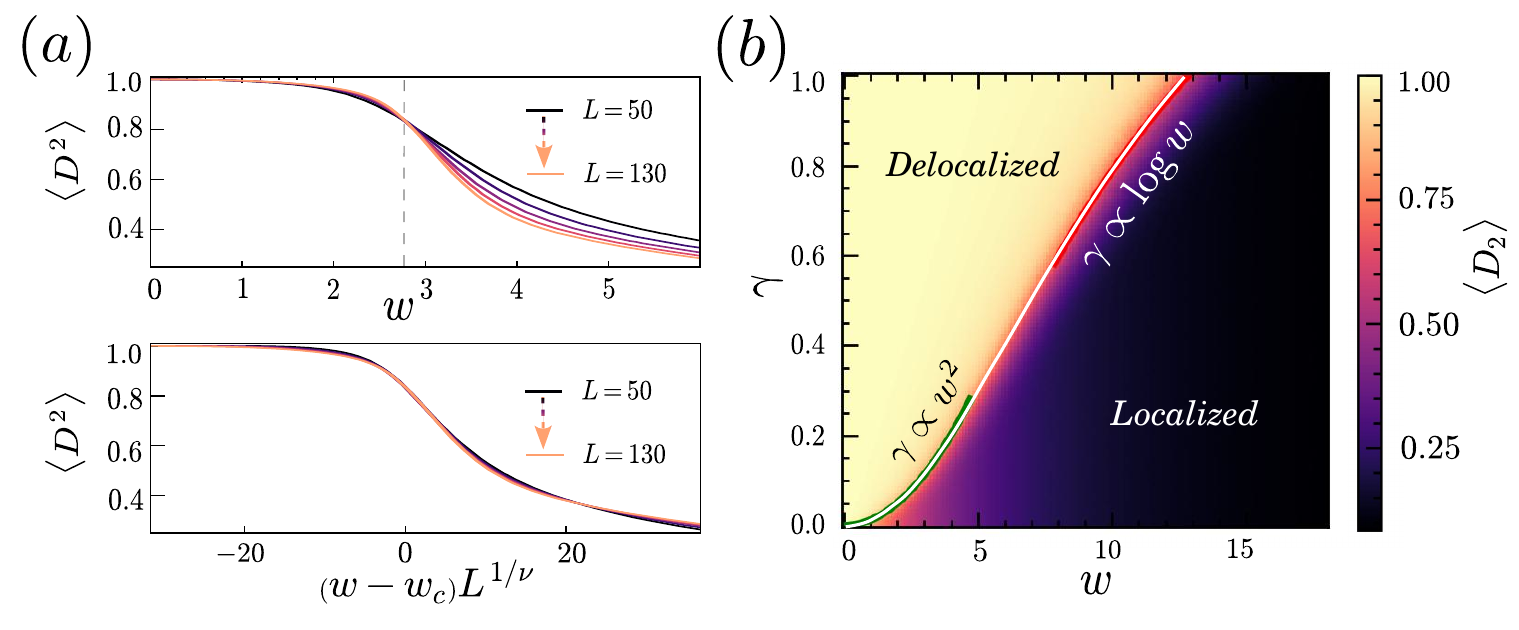}
    \caption{$\langle D_2 \rangle$ plotted as a function of ($a$) Upper panel: $w$; lower panel: $(w-w_c)L^{1/\nu}$ for $L=50,70,90,110,130$. ($b$) $\langle D_2 \rangle$ plotted in $w-\gamma$ parameter space for $L=110$.}
    \label{fig:figure_D2}
\end{figure}

The biorthogonal IPR \cite{PhysRevLett.121.026808,PhysRevB.105.075128,PhysRevB.106.064208} of the n-th eigenstate of the Hamiltonian $\mathcal{H}$ is,
\begin{align}
\mathrm{IPR\left(\mathcal{H}\right)} 
&= \frac{\sum_{j}|\langle \psi_L^{n} | j \rangle\langle j | \psi_R^{n}\rangle|^2}
{\left( \sum_{j}|\langle \psi_L^{n} | j \rangle\langle j | \psi_R^{n}\rangle|\right)^{2}}\nonumber\\
&= \frac{\sum_{j}|\langle \phi_L^{n} |\hat{S}^{-1}| j \rangle\langle j| \hat{S}| \phi_R^{n}\rangle|^2}
{\left( \sum_{j}|\langle \phi_L^{n} |\hat{S}^{-1}| j \rangle\langle j| \hat{S}| \phi_R^{n}\rangle|\right)^{2}}\nonumber \\
& = \frac{\sum_{j}|\langle \phi_L^{n} | j \rangle\langle j| \phi_R^{n}\rangle|^2}
{\left(\sum_{j}|\langle \phi_L^{n}| j \rangle\langle j| \phi_R^{n}\rangle|\right)^{2}}\left[\because \,\, S^{-1}|j\rangle\langle j| S = |j\rangle \langle j|\right]\nonumber \\
&= \mathrm{IPR\left(H_{HN}\right)}
\end{align}
Since IPR is invariant for all the iso-spectral models so fractal dimension, defined as $D_2 = -\log{(\mathrm{IPR})}/\log{L}$ is also remains invariant.\\

\prlsection{Ciritical point analysis from $D_2$} While the change in $f_c$ and $\mathcal{W}$ already indicates the presence of a phase transition (see in the main text), this spectral criterion alone does not reveal how the underlying eigenstates reorganize across the transition. IPR for the n-th eigenstate defined as, $\mathrm{IPR} = 
(\sum_{j}|\langle \psi_L^{n} | j \rangle\langle j | \psi_R^{n}\rangle|^2)/(\sum_{j}|\langle \psi_L^{n} | j \rangle\langle j | \psi_R^{n}\rangle|)^{2}$ where, $|\psi_{L}\rangle$ and $|\psi_{R}\rangle$ are the left and right eigenstates of the hamiltonian with eigenvalue equations $\mathcal{H}|\psi_{R}^n\rangle = E_n |\psi_{R}^n\rangle$ and $\langle \psi_{L}^{n}|\mathcal{H} = \langle \psi_{L}^{n}|E_n$ which satisfy the bi-orthogonal condition, $\langle \psi_L^{n}|\psi_R^m\rangle = \delta_{m,n}$. \\

Criticality is characterized by plotting $\langle D_2 \rangle$, defined as $\langle D_2 \rangle = -\log{\langle \mathrm{IPR} \rangle}/\log{L}$ \cite{PhysRevB.106.094204,RevModPhys.80.1355,HENTSCHEL1983435,PhysRevLett.50.346}, as a function of $w$ for different system sizes $L$ [Fig.\ref{fig:figure_D2}($a$)(top)]. This is complemented by a scaling collapse using the variable $(w-w_c^{(2)})L^{1/\nu}$ with $\nu = 2.0$ and $w_c^{(2)} = 2.8$ for $\gamma = 0.1$ [Fig.\ref{fig:figure_D2}($a$)(bottom)]. Here, we consider the central $20\%$ of the modes, as the modes near $\Re(E)_{\min}$ and $\Re(E)_{\max}$, corresponding to the spectral edges, are least sensitive to $\gamma$ (shown in red circles in Fig.\ref{fig:ipr_fidel}($a,b$)). The phase boundary in full parameter space is captured by tracking $(w,\gamma)$ coordinate where $\langle D_2 \rangle$ exhibit no scaling with $L$ [Fig.\ref{fig:figure_D2}(b)]. 

\subsection{Fidelity under a Local Perturbation and Its Invariance under Similarity Transformation}
To quantify the response of $\mathcal{H}$ to a strictly local perturbation, we evaluate the bi-orthogonal fidelity \cite{Tu2023generalpropertiesof,PhysRevResearch.3.013015,sun2022biorthogonal,PhysRevA.98.052116} between the many-body eigenstates of the unperturbed and perturbed Hamiltonians. The perturbation is introduced at site $1$ through a local potential $v_0$, such that
\begin{equation}
    H^{I}=\mathcal{H}, \qquad
    H^{F}= \mathcal{H} + v_0 \, n_1,
\end{equation}
where $\hat{n}_1$ is the local number operator and $v_0$ denotes the perturbation strength. The fidelity is defined as
\begin{equation}
F(v_0)=
\sqrt{
\langle \Psi_L^{I} | \Psi_R^{F} \rangle
\langle \Psi_L^{F} | \Psi_R^{I} \rangle
},
\end{equation}

where $|\Psi_R^{I,F}\rangle$ and $\langle \Psi_L^{I,F}|$ are the right and left half-filled many-body eigenstates of $H^{I}$ and $H^{F}$, respectively. We now show that the fidelity remains invariant under the similarity transformation relating the HN Hamiltonian to its isospectral counterpart.
\begin{equation}
    H^{I}_{HN}=S^{-1} H^{I} S,
\end{equation}
Since similarity transformations preserve the spectrum, $H^{I}$ and $H^{I}_{HN}$ are isospectral. Applying the same local perturbation to $H^{I}_{HN}$ yields
\begin{align}
    H^{F}_{HN} &= H^{I}_{HN} + v_0 \, n_1 \nonumber\\
    &= H^{I}_{HN} + v_0 S^{-1} n_1 S = S^{-1} H^{F} S,
\end{align}

which is therefore isospectral to $H^{F}$. \\

Denoting the corresponding left and right many-body eigenstates of $H^{I}_{HN}$ and $H^{F}_{HN}$ as $\lvert\Phi^{I}_{L,R}\rangle$ and $\lvert\Phi^{F}_{L,R}\rangle$ respectively. Using the transformation properties of the left and right eigenstates under $S$,
\begin{equation}
    |\Psi_R\rangle = S |\Phi_R\rangle, \,\, \langle \Psi_L| = \langle \Phi_L| S^{-1},
\end{equation}
one obtains
\begin{align}
F&=
\sqrt{
\langle \Psi_L^{I} | \Psi_R^{F} \rangle
\langle \Psi_L^{F} | \Psi_R^{I} \rangle
}
\nonumber\\
&=
\sqrt{
\langle \Phi_L^{I} |  S^{-1} S | \Phi_R^{F} \rangle
\langle \Phi_L^{F} |  S^{-1} S | \Phi_R^{I} \rangle
}
\nonumber\\
&=
\sqrt{
\langle \Phi_L^{I} | \Phi_R^{F} \rangle
\langle \Phi_L^{F} | \Phi_R^{I} \rangle
}
= F_{HN}.
\end{align}

Therefore, the biorthogonal fidelity is exactly invariant under similarity transformations. So the sensitivity to local perturbations is not specific to a particular representation of the Hamiltonian but is shared by all members of the corresponding isospectral class.

\subsection{Wavepacket dynamics of the HN and SBN models for $w=0$}

\begin{figure}[tbh!]
\centering
\includegraphics[width=1.0\linewidth]{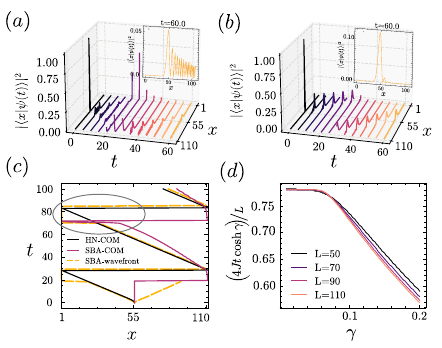}
\caption{Time evolution of an initially localized wavepacket prepared at the center of the lattice, $\psi(x)=\delta_{x,L/2}$, for a system of size $L=110$. ($a$) Dynamics in the SBN model and ($b$) dynamics in the HN model. The insets show magnified views of the probability density profile at $t=60$. ($c$) Time evolution of the COM for the HN and SBN models, together with the propagation of the wavefront in the SBN model. ($d$) Dependence of $4Jt \cosh\gamma/L$ on $\gamma$ for various system sizes of the SBN model, with the wavepacket initially prepared at a distance of $L/2$ from the non-Hermitian (nH) bond.}
\label{fig:figure5}
\end{figure}

Since the non-Hermiticity in the HN model is distributed throughout the entire lattice, a localized wavepacket immediately experiences its effect and consequently exhibits a chiral flow, as shown in Fig.~\ref{fig:figure5}($b$). In contrast, in the SBN model the wavepacket is initialized at the center of the lattice ($L/2$), far from the non-Hermitian bond. Therefore, at early times the wavepacket undergoes symmetric ballistic spreading before becoming influenced by the non-Hermitian bond. Once the wavepacket receives a signal from the non-Hermitian bond, its spreading is temporarily arrested, leading to a transient localization regime. Subsequently, the wavepacket resumes its propagation but develops a long asymmetric tail, as illustrated in Fig.~\ref{fig:figure5}($a$).

The insets of Fig.~\ref{fig:figure5}($a$) and ($b$) show the probability density profile at $t=60$. A striking feature of the SBN model is that the long tail of the wavepacket remains connected to the non-Hermitian bond throughout the evolution. By contrast, the wavepacket in the HN model exhibits a globally asymmetric profile due to the presence of non-Hermiticity at every lattice site.

Figure~\ref{fig:figure5}($c$) shows the time evolution of the COM \cite{PhysRevLett.80.1800,PhysRevB.105.024303,wsmq-kmq9,manda2026crossovers,10.21468/SciPostPhys.16.5.120} for both models. In the HN model, the COM starts moving immediately after the wavepacket is created. In the SBN model, however, the COM remains nearly stationary during the initial stage of symmetric ballistic expansion. The wavefront requires a characteristic time
\begin{equation}
t_{\mathrm{wf}}=\frac{L/2}{2J\cosh\gamma}
\end{equation}
to reach the non-Hermitian bond. Surprisingly, we observe that the non-Hermitian bond begins to influence the COM even before the wavefront reaches it. As highlighted by the gray elliptical region in Fig.~\ref{fig:figure5}($c$), the COM is pulled toward the non-Hermitian bond from a considerable distance, indicating a nonlocal influence of the localized non-Hermiticity on the wavepacket dynamics.

To quantify this effect, in Fig.~\ref{fig:figure5}($d$) we plot the dimensionless quantity $(4Jt\cosh\gamma)/L$ as a function of $\gamma$, where $t$ denotes the time at which the COM first exhibits a noticeable deviation from its initial value. We find that for weak non-Hermiticity ($\gamma\leq0.06$), this quantity remains approximately constant, implying that the wavepacket propagates nearly the same fraction of the system size before being influenced by the non-Hermitian bond. For stronger non-Hermiticity ($\gamma>0.06$), the ratio decreases approximately linearly with $\gamma$, demonstrating that the influence of the non-Hermitian bond becomes increasingly long-ranged as the non-Hermitian strength is increased.
\subsection{Lieb-Robinson Bound for SBN}
Out-of-Time-Ordered Correlator (OTOC) defined as,
\begin{equation}
    \hat{C}(t) \propto \langle\left[W(t),V(0)\right]^2 \rangle \leq || \left[W(t),V(0)\right] ||^2
\end{equation}
where,  the Hermitian  operator $\quad W(t)=e^{iH^{\dagger}t}W(x)e^{-iHt}$.\\

By using Cauchy–Schwarz inequality,
\begin{align}
&||\left[W(t),V(0)\right]||\nonumber\\
&\leq  2||W_t||\,||V(0)||\nonumber \\
& \leq 2 ||e^{iH^{\dagger}t}||\, ||e^{-iHt}||\, ||W(x)||\,||V(0)||\nonumber \\
& \leq 2\, \Big|\Big|\sum_{n=1}^{\infty} \frac{\left(iH^{\dagger}t\right)^n}{n!} \Big|\Big| \, \Big|\Big|\sum_{m=1}^{\infty} \frac{\left(-iHt\right)^m}{m!} \Big|\Big|\, ||W(x)||\, ||V(0)||\nonumber\\
&\leq \sum_{n,m=1}^{\infty} \frac{(it)^{n+m}}{n!\, m!} ||H^{\dagger}||^n\, ||H||^m \, ||W(x)||\, ||V(0)||
\end{align}

According to Frobenius norm of H is, 
\begin{align}
||H|| &= J\sqrt{2(L-2) + e^{2L\gamma}+ e^{-2L\gamma}}\nonumber\\
&= J\sqrt{2(L-3) + (e^{L\gamma}+e^{-L\gamma})^2}\nonumber \\
&\approx 2J\cosh{\left(L\gamma\right)} \, \text{when} \, L\to \infty
\end{align} 
\begin{align}
&\text{Thus,}\,\,||\left[W(t),V(0)\right]||\nonumber\\
&\leq \sum_{n,m=1}^{\infty} \frac{(t)^{n+m}}{n!\,m!} \left(2J \cosh{L\gamma}\right)^{n+m} ||W(x)||\, ||V(0)||\nonumber\\
&= \sum_{n}\frac{\left(2tJ \cosh{L\gamma}\right)^n}{n!} \sum_{m}\frac{\left(2tJ \cosh{L\gamma}\right)^m}{m!} ||W(x) ||\, ||V(0)||\nonumber \\
&= e^{4tJ\cosh{L\gamma}}||W(x)||\, ||V(0)||
\end{align}

The Lieb-Robinson velocity as, $v_{LR} = 4J\cosh{L\gamma}$. And, $ \hat{C}(t) \propto e^{8Jt\cosh{L\gamma}} ||W(x)||^2 ||V(0)||^2$. \\


\subsection{Acceleration of the COM of the wavepacket for SBN and HN}
In the Schr\"odinger picture, the expectation value of an operator $\hat{A}$ is defined as
\begin{align}
    \langle A(t) \rangle = \frac{\langle \psi(t)|A(0)|\psi(t) \rangle}{\langle \psi(t)|\psi(t)\rangle}\,\, \text{where,} \,\, |\psi(t)\rangle = e^{-i\mathcal{H}t}|\psi(0)\rangle
\end{align}
where the normalization factor is explicitly retained due to the nonunitary time evolution.
The acceleration of the center of mass of the wavepacket can be written as,  
\begin{align}
    \frac{d^2}{dt^2}\left\langle x(t) \right\rangle &= \left\langle \frac{d^2 x}{dt^2}\left(t\right) \right\rangle - 2 \left\langle \frac{d x}{dt}\left(t\right)\right\rangle \left\langle i\left(\mathcal{H}^{\dagger} -\mathcal{H}\right)(t) \right\rangle \nonumber  \\
    &+ 2\left\langle x(t)\right\rangle \left(\left\langle i\left(\mathcal{H}^{\dagger} - \mathcal{H}\right)(t) \right\rangle\right)^2 \nonumber \\
    &- \left\langle \hat{x}(t) \right\rangle \left\langle \left[-\mathcal{H}^{\dagger}\left(\mathcal{H}^{\dagger} -\mathcal{H}\right) + \left(\mathcal{H}^{\dagger} -\mathcal{H}\right)\mathcal{H} \right]\right\rangle
\end{align}
Decomposing the non-Hermitian Hamiltonian into its Hermitian and anti-Hermitian components, $H=\mathcal{H}_h+\mathcal{H}_{ah},\,  H^\dagger=\mathcal{H}_h-\mathcal{H}_{ah},$ one obtains
\begin{align}
    \frac{d^2}{dt^2}\left\langle x(t) \right\rangle &= \left\langle \frac{d^2 x}{dt^2}\left(t\right) \right\rangle + 4i \left\langle \frac{d x}{dt}\left(t\right)\right\rangle\left\langle \mathcal{H}_{ah}(t)\right\rangle - 8 \left\langle x(t)\right\rangle\cdot \nonumber \\ 
    &\left\langle \mathcal{H}_{ah}(t) \right\rangle^2 - 2\left\langle \hat{x}(t) \right\rangle \left\langle \left(\mathcal{H}^{\dagger}\mathcal{H}_{ah} - \mathcal{H}_{ah}H \right)(t)\right\rangle
\end{align}

In Heisenberg picture,
\begin{align}
    \hat{x}(t) &= e^{i\mathcal{H}^{\dagger}t}\hat{x}(0) e^{-i\mathcal{H}t} \nonumber \\
    \frac{d\hat{x}(t)}{dt} &= i e^{i\mathcal{H}^{\dagger}t}\left(\left[\mathcal{H}_{h},\hat{x}(0)\right] - \left\{\mathcal{H}_{ah},\hat{x}(0)\right\} \right)e^{-i\mathcal{H}t} \nonumber \\
    \frac{d^2\hat{x}(t)}{dt} &= e^{i\mathcal{H}^{\dagger}t}\left(-\left[\mathcal{H}_{h},\left[\mathcal{H}_{h},\hat{x}(0)\right]\right] -\left\{\mathcal{H}_{ah},\left\{\mathcal{H}_{ah},\hat{x}(0)\right\}\right\} \right. \nonumber \\
    &\quad \left. +\left[\mathcal{H}_{h},\left\{\mathcal{H}_{ah},\hat{x}(0)\right\}\right] + \left\{\mathcal{H}_{ah},\left[\mathcal{H}_{h},\hat{x}(0)\right]\right\} \right)e^{-i\mathcal{H}t}
\end{align}

\begin{widetext}
Substituting these expressions into the normalized expectation value yields,
\begin{align}
    \frac{d^2}{dt^2}\langle x(t) \rangle &= -\langle \left[\mathcal{H}_{h},\left[\mathcal{H}_{h},\hat{x}(0)\right]\right](t)\rangle - \langle \left\{\mathcal{H}_{ah},\left\{\mathcal{H}_{ah},\hat{x}(0)\right\}\right\}(t)\rangle + \langle \left[\mathcal{H}_{h},\left\{\mathcal{H}_{ah},\hat{x}(0)\right\}\right](t)\rangle\nonumber\\
    &+ \langle \left\{\mathcal{H}_{ah},\left[\mathcal{H}_{h},\hat{x}(0)\right]\right\}(t)\rangle -4 \langle \left[\mathcal{H}_{h},\hat{x}(0)\right](t) \rangle \langle \mathcal{H}_{ah}(t) \rangle + 4 \langle \left\{\mathcal{H}_{ah},\hat{x}(0)\right\}(t) \rangle \langle \mathcal{H}_{ah}(t) \rangle\nonumber\\
    &- 8 \left\langle x(t) \right\rangle\left\langle \mathcal{H}_{ah}(t) \right\rangle^2  -2 \langle x(t) \rangle \langle \left[\mathcal{H}_{h},\mathcal{H}_{ah}\right](t) \rangle + 8 \left\langle x(t) \right\rangle \left\langle \mathcal{H}_{ah}^2(t) \right\rangle
\end{align}

For the HN and SBN models,
\begin{align}
    H_{HN}  &= -J\cosh{\gamma}\sum_{j=1}^{L}\left(c_{j}^{\dagger}c_{j+1} + c_{j+1}^{\dagger}c_j\right) +\sum_{j=1}^{L}h_j n_j - J\sinh{\gamma}\sum_{j=1}^{L}\left(c_{j}^{\dagger}c_{j+1} - c_{j+1}^{\dagger}c_j\right)\nonumber\\
    H_{SBN} &= -J\sum_{j=1}^{L-1}\left(c_{j}^{\dagger}c_{j+1} + c_{j+1}^{\dagger}c_j\right) - J\cosh{L\gamma}\left(c_L^{\dagger}c_1 + c_1^{\dagger}c_L\right) +\sum_{j=1}^{L}h_j n_j -J\sinh{L\gamma}\left(c_{1}^{\dagger}c_{L} - c_{L}^{\dagger}c_1\right)
\end{align}

A key distinction in the oscillation of COM between the two models lies in their Hermitian sectors. The Hermitian part of the HN Hamiltonian is translationally invariant, whereas that of the SBN Hamiltonian contains a locally modified bond and is therefore spatially anisotropic. As a consequence, the COM dynamics in the SBN model acquires an additional contribution originating from the anisotropic Hermitian bond. \\
In particular,
\begin{equation}
\left[H_{h},\left[H_{h},\hat{x}\right]\right] =
\begin{cases}
0 & \text{for HN} \\
LJ^2\left(\cosh{L\gamma}-1\right)\left[\left(c_{L}^{\dagger}c_2 - c_{L-1}^{\dagger}c_{1} + h.c.\right) + 2\left\{1-(L-1)\cosh{L\gamma}\right\}\left(\hat{n}_1 - \hat{n}_L\right) \right]& \text{for SBN}
\end{cases}
\end{equation}
creates oscillation in acceleration of COM of the wavepacket for SBN model. Because of the terms $(\hat{n}_1 - \hat{n}_L)$ and $(c_L^{\dagger}c_2 - c_{L-1}^{\dagger}c_1 +h.c.)$ wavepacket in SBN model shows oscillation. [Note: We numerically checked that the other terms also contribute to the oscillation, but doesn't change the overall structure.] \\

\begin{figure*}[tbh]
    \centering
    \includegraphics[width=0.95\linewidth]{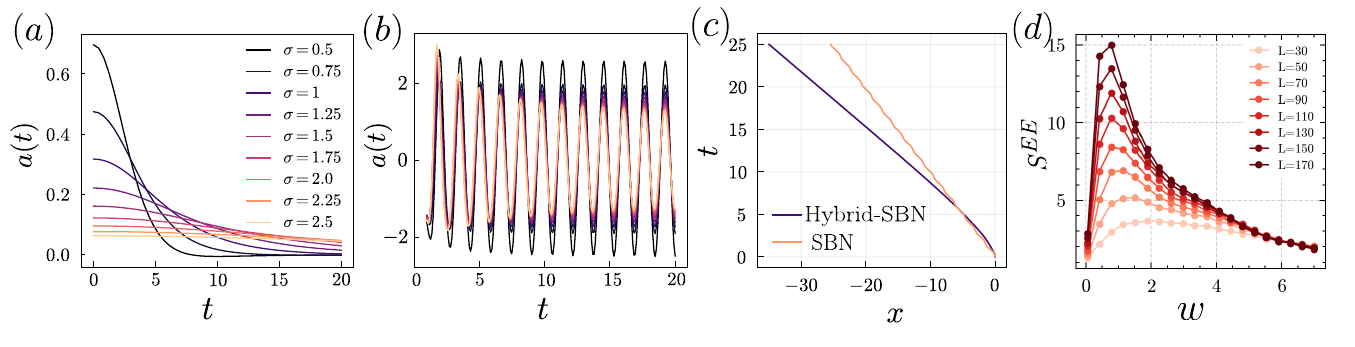}
    \caption{Short time evolution of $a(t)$ for a Gaussian wave packet in the ($a$) HN and ($b$) SBN models with $\gamma = 0.1$ and $w = 0$, plotted for spatial standard deviations ranging from $\sigma=0.5$ to $\sigma=2.5$ in steps of $0.25$. ($c$) COM motion of an initially localized wave packet at site 1 for the SBN and hybrid SBN models under the same conditions. ($d$) EE for many particle wavefunction at steady state limit, considering $A\in [0,L/2]$ and rest is subsystem $B$. So non hermitian bond resides at the bi-partition boundary.}
    \label{fig:gaussian_wavepacket}
\end{figure*}

To understand the characteristic oscillation frequency, we expand an initially localized wave packet as
\begin{align}
|\psi(0)\rangle_R = a|E_0\rangle_R +b|E_\pi\rangle_R +\sum_k c_k|E_k\rangle_R, \quad
{}_L\langle\psi(0)| = a^*{}_L\langle E_0| +b^*{}_L\langle E_\pi| +\sum_k c_k^*{}_L\langle E_k|.
\end{align}

\end{widetext}

The states $|E_0\rangle$ and $|E_\pi\rangle$ are separated from the rest of the spectrum by the largest energy scale $\Re{(E_0)}=-2J\cosh\gamma,\,\, \Re{(E_\pi)}=2J\cosh\gamma.$
The corresponding time-evolved state is
\begin{align}
|\psi(t)\rangle = a\,e^{-iE_0t}|E_0\rangle_R
&+b\,e^{-iE_\pi t}|E_\pi\rangle_R \nonumber  \\
&+\sum_k c_k e^{-iE_k t}|E_k\rangle_R,
\end{align}
with an analogous expression for the left state. \\

At short times, when many eigenstates contribute comparably to the dynamics, the dominant oscillation frequency is set by the energy separation between the extremal states,
\begin{align}
\omega = |E_\pi-E_0| = 4J\cosh\gamma .
\end{align}

At longer times, the dynamics becomes increasingly dominated by eigenstates with the largest imaginary parts of the spectrum. Consequently, the oscillation frequency decreases and eventually vanishes once the wave-packet propagation ceases.\\

And this oscillatory acceleration is not restricted to a localized initial state in SBN. Similar behavior is observed for an initial Gaussian wave packet, $\lvert \psi(0) \rangle=\left(1/\sqrt{2\pi\sigma^2}\right)\exp{\left(-x^2/(2\sigma^2)\right)}$ (Fig.\ref{fig:gaussian_wavepacket}($b$)). For HN the gaussian wavepacket have non-zero acceleration for long time for large $\sigma$ but for dies almost immediately for small $\sigma$ (Fig.~\ref{fig:gaussian_wavepacket}(a)).\\

The physical origin of the oscillations becomes particularly transparent in a hybrid SBN model defined by
\begin{align}
H_h^{\rm hybrid}
&= -J\sum_{j=1}^{L} \left( c_j^\dagger c_{j+1}
+c_{j+1}^\dagger c_j \right),\\
H_{ah}^{\rm hybrid}
&=-J\sinh(L\gamma)\left(c_1^\dagger c_L-c_L^\dagger c_1\right).
\end{align}

Here translational invariance is restored in the Hermitian sector while retaining the non-Hermitian boundary bond. We find that the oscillations in the COM acceleration disappear completely. Moreover, the COM propagates significantly faster than in the original SBN model [Fig.~\ref{fig:gaussian_wavepacket}(c)]. This demonstrates that the oscillatory acceleration originates from the anisotropic Hermitian bond rather than from the nonreciprocal hopping itself. Once translational invariance is restored, the wave-packet tail no longer remains effectively coupled to the non-hermitian bond, leading to the suppression of the oscillatory behavior.\\

\subsection{SSEE for different subsystem choice}
Dip (or second reemergence) in the steady state entanglement entropy (SSEE) only visible when nH bond is kept away from subsystem boundary (away from boundary of $A\in\{L/4,L/4+1,\dots,3L/4\}$ and rest is $B$).\\

To prove this we choose $A\in\{0,1,\dots,L/4\}$ and rest is subsystem $B$. So the nH band is at the boundary of the two subsystems. And we observe that there is no dip (no reemergence) in SSEE (see Fig.~\ref{fig:gaussian_wavepacket}(d)). 

\subsection{GKSL equation and jump operators for isospectral models}
Equation [\ref{eq:sm_eq1}] can be break into hermitian and anti-hermitian part as, $\mathcal{H} = \mathcal{H}_{h} + \mathcal{H}_{nh}$, where,
\begin{align*}
\mathcal{H}_{h} &= -J \sum_{j=1}^{L-1}\cosh{(\gamma-g_j})\left(c_j^{\dagger}\,c_{j+1} + h.c.\right)  + \sum_{j=1}^{L}h_{j}\,c_{j}^{\dagger}\,c_{j},\quad\\
\mathcal{H}_{nh} &= -J\sum_{j=1}^{L}\sinh{(\gamma-g_j)}\left(c_{j}^{\dagger}\,c_{j+1} - c_{j+1}^{\dagger}\,c_{j}\right) 
\end{align*}
Now, according to Lindblad master equation \cite{Manzano_2020,10.1093/acprof:oso/9780199213900.001.0001}, 
\begin{align*}
\frac{d\rho}{dt} & = -i \left[\mathcal{H}_{h},\rho\right]+\sum_{\mu}L_{\mu}\rho L_{\mu}^{\dagger} - \frac{1}{2}\sum_{\mu}\left\{L_{\mu}^{\dagger}L_{\mu},\rho\right\} \\
& = -i\left(\mathcal{H}_{eff}\,\rho - \rho\,\mathcal{H}_{eff}^{\dagger}\right) +\sum_{\mu}L_{\mu}\rho L_{\mu}^{\dagger}
\end{align*}

where, $\mathcal{H}_{eff} =  \mathcal{H}_{h}-\frac{i}{2}\sum_{\mu}L_{\mu}^{\dagger}L_{\mu}$\\

With the local jump operators,
\begin{align*}
    L_{j}^{l} &= \sqrt{J\sinh(\gamma-g_j)}\left(c_j - i c_{j+1}\right), \\
    L_{j}^{g} &= \sqrt{J\sinh(\gamma-g_j)}\left(c_j^{\dagger} - i c_{j+1}^{\dagger}\right),
\end{align*} 
one can get, $\mathcal{H}_{eff} = \mathcal{H} - iJ\sum_{j=1}^{L}\sinh{\left(\gamma-g_j\right)}$. Extra complex energy term $-J\sum_{j=1}^{L}\sinh{\left(\gamma-g_j\right)}$, which only shifts the energy spectrum in the imaginary direction and doesn't change any physics we are interested in this paper \cite{PhysRevLett.68.580,carmichael1993open,Chru_ci_ski_2022,PhysRevA.108.032214}. \\

\centering
\noindent\rule{0.30\textwidth}{0.4pt}

\end{document}